\documentclass[12pt]{iopart}

\usepackage{iopams}
\usepackage{graphicx}                             
\usepackage{color}

\begin{document}

\title[Disorder in screening theory]{The effect of disorder within the interaction theory of integer quantized Hall effect}

\author{S. E. Gulebaglan$^1$, G. Oylumluoglu$^2$, U. Erkarslan$^2$,\\ A. Siddiki$^{2,3}$ and I. Sokmen$^1$}
\address{$^1$Dokuz Eyl\"ul University, Physics Department, T\i naztepe Campus, 35100 \.Izmir, Turkey}
\address{$^2$Mu\~gla University, Physics Department, Faculty
of Arts and Sciences, 48170-K\"otekli, Mu\~gla, Turkey}
\address{$^3$Istanbul University, Faculty
of Sciences, Physics Department, Vezneciler-Istanbul 34134, Turkey}
\ead{afifsiddiki@gmail.com}
\begin{abstract}
We study effects of disorder on the integer quantized Hall effect within
the screening theory, systematically. The disorder potential is
analyzed considering the range of the potential fluctuations. Short range
part of the single impurity potential is used to define the
conductivity tensor elements within the self-consistent Born
approximation, whereas the long range part is treated
self-consistently at the Hartree level. Using the simple, however, fundamental
Thomas-Fermi screening, we find that the long range disorder
potential is well screened. While, the short range part is
approximately unaffected by screening and is suitable to define
the mobility at vanishing magnetic fields. In light of
these range dependencies we discuss the extend of the quantized
Hall plateaus considering \emph{the} ``mobility" of the wafer and
the width of the sample, by re-formulating the Ohm's law at low
temperatures and high magnetic fields. We find that, the plateau
widths mainly depend on the long range fluctuations of the
disorder, whereas the importance of density of states broadening
is less pronounced and even is predominantly suppressed. These
results are in strong contrast with the conventional single
particle pictures. We show that the widths of the quantized
Hall plateaus increase with increasing disorder, whereas the level broadening is negligible.

\end{abstract}
This work focuses on the disorder effects on the integer quantized Hall effect within the screening theory. Since the early days of QHE, disorder played a very important role, however, interactions were completely neglected. Here we present our results which also includes interactions in a self-consistent manner and show that even without localization one can obtain the quantized Hall plateaus. We investigated different aspects of the impurity potential and suggested a criterion on mobility at high magnetic fields. We think that our work will shed light on the understanding of the QHE and is interest to condensed matter community.
\vspace{2pc}
\submitto{\JPC}
\maketitle

\section{Introduction\label{intro}}
The integer quantized Hall effect (IQHE), observed at two
dimensional charge systems (2DCS) subject to strong perpendicular
magnetic fields $B$, is usually discussed within the single
particle picture, which relies on the fact that the system is
highly disordered~\cite{Kramer03:172,Schweitzer85:379}. These quantized
(spinnless) single particle energy levels are called the Landau
levels (LLs) and the discrete energy values are given by
$E_N=\hbar\omega_c(n+1/2)$, where $n$ is the Landau index and
$\omega_c=eB/m^*c$ is the cyclotron frequency of an electron with
an effective mass $m^*$ ($\approx 0.067m_e$, $m_e$ being the
bare electron mass at rest) and $c$ is the speed of light in
vacuum. In single particle models the disorder plays several
roles, such as Landau level broadening~\cite{Cai86:3967}, leading
to a finite longitudinal
conductivity~\cite{Ando74:959,Ando75:279}, spatial
localization~\cite{Nixon:90} \emph{etc}. Disorder can be created
by inhomogeneous distribution of dopant ions which essentially
generates the confinement potential~\cite{stopadisorder:96} for
the electrons. In the absence of disorder, the
density of states are Dirac delta-functions
$D(E)=\frac{1}{2 \pi
l^2}{\sum^{\infty}_{N=0}}{\delta(E-E_N)}$, where
$l=\sqrt{\hbar/eB}$ is the magnetic length, and the longitudinal
conductivity ($\sigma_{l}$) vanishes. For a homogeneous two dimensional electron system (2DES), by
the inclusion of disorder and due to collisions, LLs become
broadened. Therefore the longitudinal conductance becomes non-zero
in a finite energy (in fact magnetic field) interval. Long range
potential fluctuations generated by the disorder result in the so
called {\em classical localization}~\cite{Fogler94:1656},
\emph{i.e.} the guiding center of the cyclotron orbit moves along
closed equi--potentials~\cite{Efros88:1019}. In contrast to the
above mentioned bulk theories, the edge theories usually disregard
the effect of disorder to explain the (quantized) Hall resistance
$R_{\rm H}$ and accompanying (zero) longitudinal resistance
$R_{\rm L}$. However, the non-interacting edge theories still
require disorder to provide a reasonable description of the
transition between the plateaus. The Landauer-B\"uttiker approach
(known as the edge channel picture) \cite{Buettiker86:1761} and
its direct Coulomb interaction generalized version, \emph{i.e.}
the non-self-consistent Chklovskii
picture~\cite{Chklovskii92:4026}, also needs localization
assumptions in order to obtain quantized Hall (QH) plateaus of
finite width (see for a review \emph{e.g.} Datta's book
\cite{Datta} and Ref.~\cite{Efros88:1019} for the estimates
of plateau widths at the high disorder limit).

In contrast to above discussions very recent
experimental~\cite{Wilde06:disorder,Mares:09,josePHYSE,jose:prl}
and theoretical~\cite{dassarma:mobility,Macleod09:background}
results point the incomplete treatment of the disorder potential
and scattering mechanisms. Fairly recent theoretical
approaches~\cite{Guven03:115327}, the QH plateaus are obtained by
the inclusion of direct Coulomb interaction
self-consistently~\cite{siddiki2004} and the effect of the
disorder was handled via conductivity tensor
elements~\cite{Bilayersiddiki06:}, however, the source of the
disorder and its properties was left
unresolved~\cite{Siddiki04:condmat}. Whereas, the influence of
potential fluctuations on the QH plateaus were discussed
briefly~\cite{Gerhardts08:378,Siddiki:ijmp}.

This work provides a systematic investigation of the disorder
potential and its influence on the quantized Hall effect including
direct Coulomb interaction. The investigation is extended to realistic experimental
conditions in determining the widths of the quantized Hall
plateaus. We, essentially study the effect of disorder in two
distinct regimes, namely the short range and the long range. The
short range part is included to the density of states (DOS),
thereby influences the widths of the current carrying edge-states
and the entries of the conductivity tensor. Whereas, the long
range part is incorporated to the self-consistent calculations,
which determines the extend of the plateaus in turn. In
Sec~\ref{impurity} we introduce two types of single
impurity potentials, namely the Coulomb and the Gaussian, and
compare their range dependencies considering damping of the
dielectric material. In the next step we
discuss the \emph{screened} disorder potential within a pure electrostatic approach, by considering an
homogeneous two dimensional electron system (2DES) without an
external magnetic field and show that the long range part is well
screened, whereas the short range part is almost unaffected. Section~\ref{treed} is devoted to investigate
impurities numerically, where we solve the Poisson equation
self-consistently in three dimensions. The numerical and analytical calculations
are compared, considering the estimations of the disorder
potential range and its variation amplitude. We finalize our
discussion with Sec.~\ref{quanHP}, where we calculate the plateau
widths under experimental conditions for different sample widths
and mobilities.
\section{Impurity potential\label{impurity}}
The disorder potential experienced by the 2DES, resulting from the
impurities has quite complicated range dependencies. Since, the
potential generated by an impurity is (i) \emph{damped} by the
dielectric material in between the impurity and the plane where
the 2DES resides (ii) is screened by the homogeneous 2DES
depending on the density of states, which changes drastically with
and without magnetic field. It is common to theoreticians to
calculate the conductivities from single impurity potentials, such
as Lorentzian~\cite{Guven03:115327}, Gaussian~\cite{Ando82:437} or
any other analytical functions~\cite{Champel08:124302,TobiasK06:h}. However, the landscape of potential fluctuations
is also important to define the actual mobility of the sample at
hand, in particular in the presence of an external magnetic field.
\subsection{Pure Electrostatics}
We first discuss the different range dependencies
of the Coulomb and Gaussian donors, assuming open boundary conditions. Next, the effect of
the spacer thickness on the disorder potential is discussed, namely
the damping of the external (Coulomb) potential, and is compared
with the Thomas-Fermi screening. The different damping/screening
dependencies of the resulting potentials are discussed in terms of
range.

The Coulomb potential presents long range part, which leads to long range fluctuations due to
overlapping if several donors are considered. Whereas, the Gaussian potential decays exponentially on the length
scale comparable with the separation thickness. Since the Gaussian potential is relatively
short ranged, no overlapping of the
single donor potentials occur. Hence, the external potential
experienced by the electrons can be approximated to a homogeneous
potential fairly good. Thus one can conclude that approximating the
total disorder potential by Gaussians is not sufficient to
recover the long range part. Similar
arguments are also found in the
literature~\cite{Nixon:90,Efros88:1019,Siddiki:ijmp}. In order to
overcome the difference observed at the long range potential
fluctuations between the Coulomb and the Gaussian impurities, the
following procedure is applied: First we calculate the total disorder potential considering many impurities then we perform a two-dimensional
Fourier transformation of the Coulomb potential and make a back
transformation keeping the first few momentum $q$ components in
each direction, hence only the long range part of the potential is
left~\cite{Siddiki:ijmp}. Then we add the long range part of the Coulomb potential to
the potential created by donors, \emph{i.e.} to the confinement
potential. We take this as a motivation to simulate the short
range part of the impurity potential by Gaussian impurities, and
calculate the Landau level broadening and the conductivities,
described within the self-consistent Born
approximation (SCBA)~\cite{Ando82:437}.

Here we point to the effect of the spacer thickness on the impurity
potential experienced in the plane of 2DES. It is well known from
experimental and theoretical investigations that, if the distance
between the electrons and donors is large, the mobility is
relatively high and it is usually related with suppression of the
short range fluctuations of the disorder potential. These
results agree with the experimental observations of high mobility
samples and are easy to understand from the $z$ dependence of the
Fourier expansion of the Coulomb potential, \begin{equation}V^{
\vec{q}}(z)=\int d \vec{r} e^{-i \vec{q} \cdot \vec{r}}
\sum_{j}^{N} \frac{e^{2}/ \bar{\kappa}}{ \sqrt{( \vec{r}-
\vec{r_{j}})^{2}+z^{2}}}=\frac{2 \pi
e^{2}}{\bar{\kappa}q}e^{-|qz|}NS(\vec{q}), \label{fourier1}\end{equation}where $S(\vec{q})$ contains all the information about the in-plane
donor distribution and $N$ is the total number of the ionized
donors. We observe that if the spacer thickness is increased, the
amplitude of the potential decreases rapidly. We also see that the
short range potential fluctuations, which correspond to higher
order Fourier components, are suppressed more efficiently.

Next, we discuss electronic screening of the external potential
created by the donors discussed above. For a dielectric material
the relation between the external and the screened potentials are
given by, \begin{equation}V_{\rm scr}^q=V_{\rm ext}^q/\epsilon(q),
\label{linear-eps}\end{equation}where $\epsilon(q)$ is the dielectric
\emph{function} and is given by $\epsilon(q)=1+\frac{2\pi
e^2D_0}{\bar{\kappa}|q|}, $ with the constant 2D density of
states $D_0=m/(\pi \hbar^2)$ in the absence of an external $B$
field, and is known as the Thomas-Fermi (TF) function.
The simple linear relation above, together with the TF dielectric
function essentially describes the electronic screening of the
Coulomb potential given in Eq.~\ref{fourier1}, if there are
sufficient number of electrons~\cite{Efros88:1019} ($n_{\rm
el}>0.1\cdot10^{15}$ m$^{-2}$). Consider a case where the $q$
component approaches to zero, then the external (damped) potential
is well screened, hence the long range part of the disorder
potential. Whereas, the short range part remain unaffected,
\emph{i.e.} high $q$ Fourier components. Now we turn our attention
to the second type of impurities considered, the Gaussian ones. As
well known, the Fourier transform of a Gaussian is also of the
form of a Gaussian, therefore, similar arguments also hold for
this kind of impurity.

We should emphasize once more the clear
distinction between the effect of the spacer on the external
potential and the screening by the 2DES, \emph{i.e.} via
$\epsilon(q)$. The former depends on the Fourier transform of the
Coulomb potential and the important effect is the different decays
of the different Fourier components (see Eq.~\ref{fourier1}), so
that the short range part of the disorder potential is well
dampened, whereas the latter depends on the relevant DOS of the
2DES and the screening is more effective for the long range part.

\begin{figure}[!t]\center{
\includegraphics[width=0.55\linewidth]{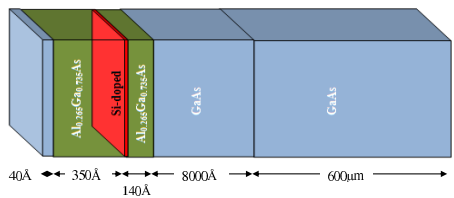}
\vspace{-7cm}
\caption{Schematic representation of the crystal, which we investigate numerically.
The crystal is grown on a thick GaAs substrate, where the 2DES is
formed at the interface of the AlGaAs/GaAs hetero-junction.
The top AlGaAs layer is doped with Silicon 30 nm above the
interface. The crystal is spanned by a 3D matrix
($128\times128\times60$). \label{fig:5}}}
\end{figure}
We continue our investigation by solving the 3D Poisson equation
iteratively for randomly distributed single impurities, where
three descriptive parameters (\emph{i.e.} the number of
impurities, the amplitude of the impurity potential and the
separation thickness) are analyzed separately. Next, we discuss the
long range parts of the potential fluctuations investigating the
Coulomb interaction of the 2DES, numerically. The range is estimated from these investigations by performing Fourier
analysis and is related to the samples used in
experiments~\cite{josePHYSE,jose:prl} (Sec.~\ref{exper}).

\subsection{3D simulations\label{treed}}
In the previous section we took a rather simple way to study the
effect of interactions by assuming an homogeneous 2DES and
screening is handled by the TF dielectric function. Here, we
present our results obtained from a rather complicated numerical
method. We solve the Poisson equation in 3D starting from the
material properties of the wafer at hand, the typical material we
consider is sketched in Fig.~\ref{fig:5}. Namely, using the growth
parameters, we construct a 3D lattice where the potential and the
charge distributions are obtained iteratively assuming open
boundary conditions, \emph{i.e.} $V(x\rightarrow \pm\infty,
y\rightarrow\pm\infty, z\rightarrow\pm\infty)=0$. For such
boundary conditions, we chose a lattice size which is considerably
larger than the region that we are interested in. We preserve the
above conditions within a good numerical accuracy (absolute error
of $10^{-6}$). A forth order grid
approach~\cite{Weichselbaum03:056707} is used to reduce the
computational time, which is successfully used to describe similar
structures~\cite{Sefa08:prb}.

Figure~\ref{fig:5} presents the schematic drawing of the
hetero-structure which we are interested in. The donor layer is
$\delta-$ doped by a density of $3.3\times10^{16}$ m$^{-2}$
(ionized) Silicon atoms, $\sim 30$ nm above the 2DES, which
provide electrons both for the potential well at the interface and
the surface. It is worthwhile to note that most of the electrons
($\sim \%90$) escape to the surface to pin the Fermi energy to the
mid-gap of the GaAs. In any case, for such wafer parameters
there are sufficient number of electrons ($n_{\rm el}\gtrsim 3.0
\times10^{15}$ m$^{-2}$) at the quantum well to form a 2DES. To
investigate the effect of impurities we place positively charged
ions at the layer where donors reside. From Eq.~\ref{fourier1} we
estimate the amplitude of the potential of a single impurity to be
$\frac{e^2}{\kappa}\frac{V_{\rm imp}}{z_D}=0.033$ eV and assume
that \emph{some} percent of the ionized donors are generating the
disorder potential, that defines the long range fluctuations. In
our simulations we perform calculations for a unit cell with
areal size of $1.5~\mu$m$\times 1.5~\mu$m which contains $3.3
\times10^{16}$ donors per square meters, thus with 10 percent
disorder we should have $N_I\sim$ 3300 impurities.
\begin{figure}[!t]\centering
\begin{minipage}{1\linewidth}
\includegraphics[width=0.5\linewidth,angle=0]{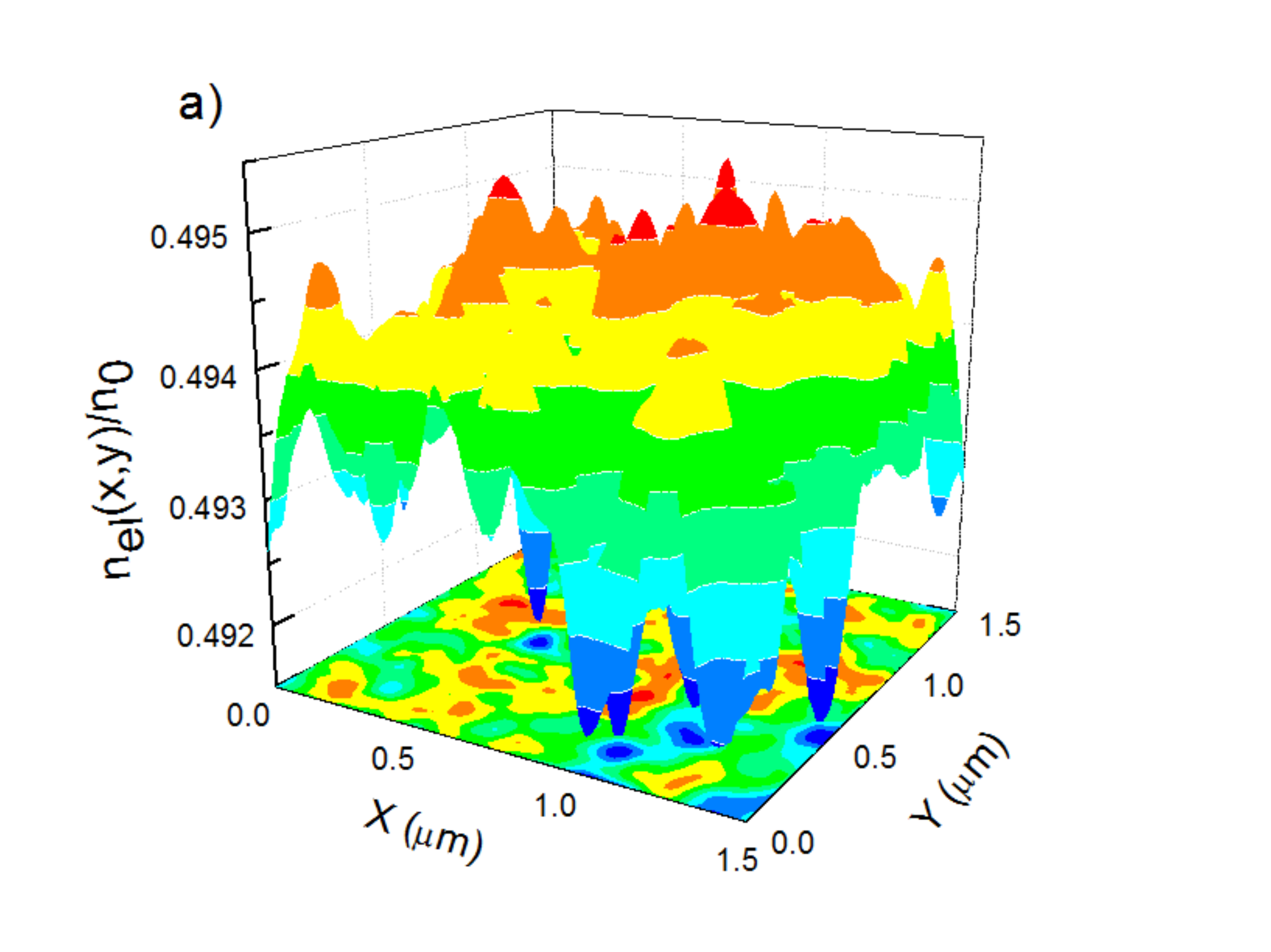}
\includegraphics[width=0.5\linewidth,angle=0]{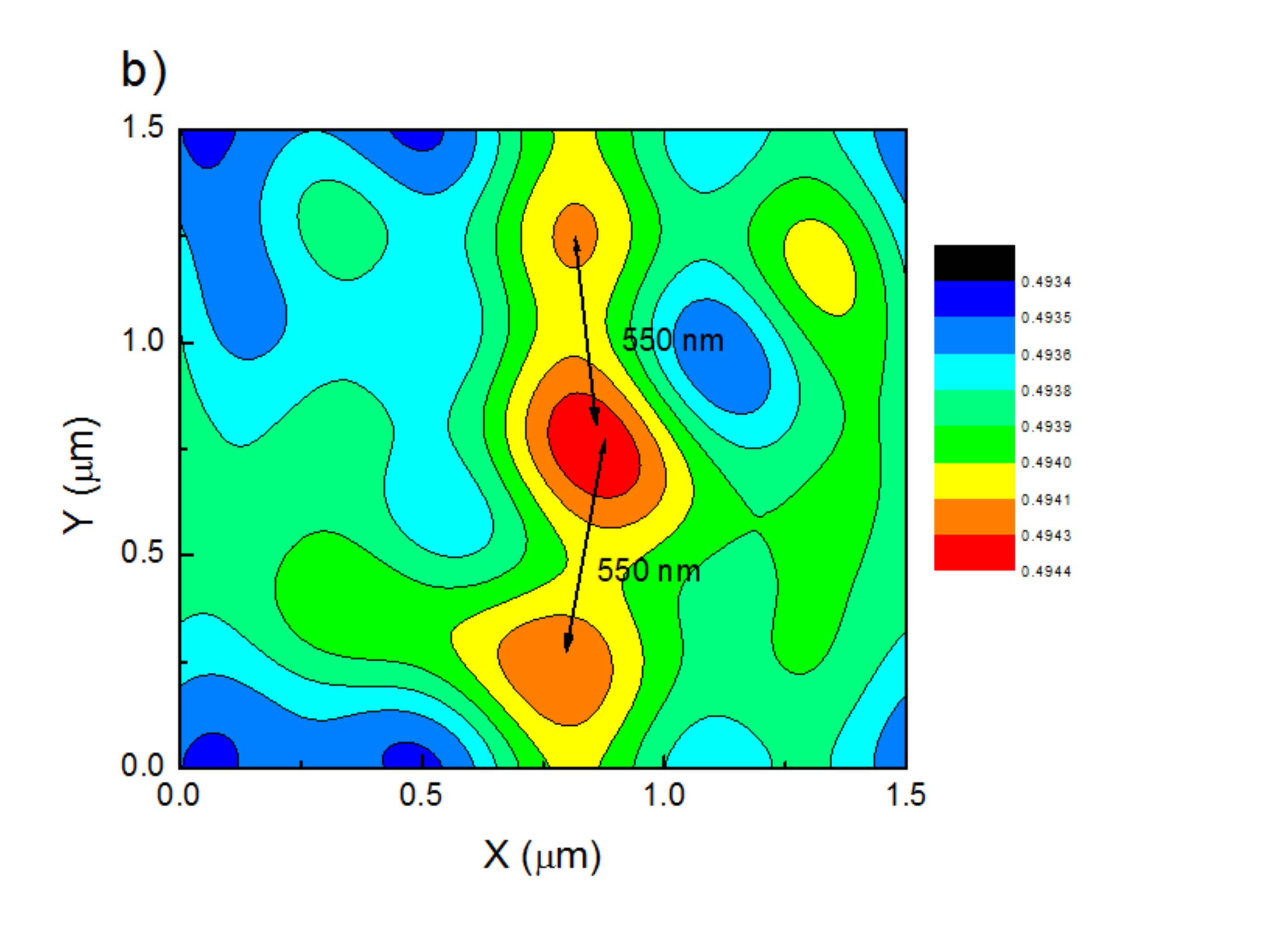}
\end{minipage}
\caption{(a) Electron density fluctuation considering 3300
impurities 30 nm above the electron gas. (b) The long-range part,
arrows are to guide the distance between two maxima. The
calculation is repeated for 50 random distributions, which lead to
a similar range.\label{fig:7}}
\end{figure}
Figure~\ref{fig:7}a depicts the actual density distribution,
when considering 3300 impurities, whereas Fig.~\ref{fig:7}b
presents only the long range part of the density fluctuation. The
arrows show the average distance between two maxima, which is
calculated approximately to be $550$ nm. To estimate an average
range of the disorder potential, we repeated calculations for such
randomly distributed impurities, where number of repetitions
scales with $\sqrt{N_I}$. Such a statistical investigation,
sufficiently ensembles the system to provide a reasonable
estimation of the long range fluctuations. We also tested for
larger number of random distributions, however, the estimation
deviated less than tens of nanometers. We show our main result of
this section in Fig.~\ref{fig:8}, where we plot the estimated long
range part of the disorder potential considering various number of
impurities $N_I$ and impurity potential amplitude $V_{\rm imp}$.
Our first observation is that the long range part of the total
potential becomes less when $N_I$ becomes large, not surprisingly.
However, the range increases nonlinearly while decreasing $N_I$,
obeying almost an inverse square law and tend to saturate at
highly disordered system. When fixing the distributions and $N_I$,
and changing the amplitude of the impurity potential we observe
that for large amplitudes the range can differ as large as 200 nm
at all impurity densities. We found that for impurity
concentration less than $\% 3$, the range of the potential is
larger than the unit cell we consider, \emph{i.e} $R>1.5 \mu$m. In
contrast to the long range part, the short range part is almost
unaffected by the impurity concentration, however, is affected by
the amplitude. Therefore, while defining the conductivities we
will focus our investigation on $V_{\rm imp}$. Another important
result is that the estimates of long range fluctuations does not
depend strongly on the spacer thickness, if one keeps the
amplitude of single impurity potential amplitude fixed,
Fig.~\ref{fig:8}b. All of the above numerical observations
coincide fairly good with our analytical investigations in the
previous section. However, the range dependency on the impurity
concentration cannot be estimated with the analytical formulas
given. We should also note that, similar or even complicated
numerical calculations are present in the
literature~\cite{Nixon:90,stopadisorder:96}. A indirect measure of
the screening effects on the potential can also be inferred by
capacitance measurements, supported by the above calculation
scheme in the presence of external field~\cite{Mares:09}.
\begin{figure}[!t]\centering
\begin{minipage}{1\linewidth}
\includegraphics[width=.5\linewidth,angle=0]{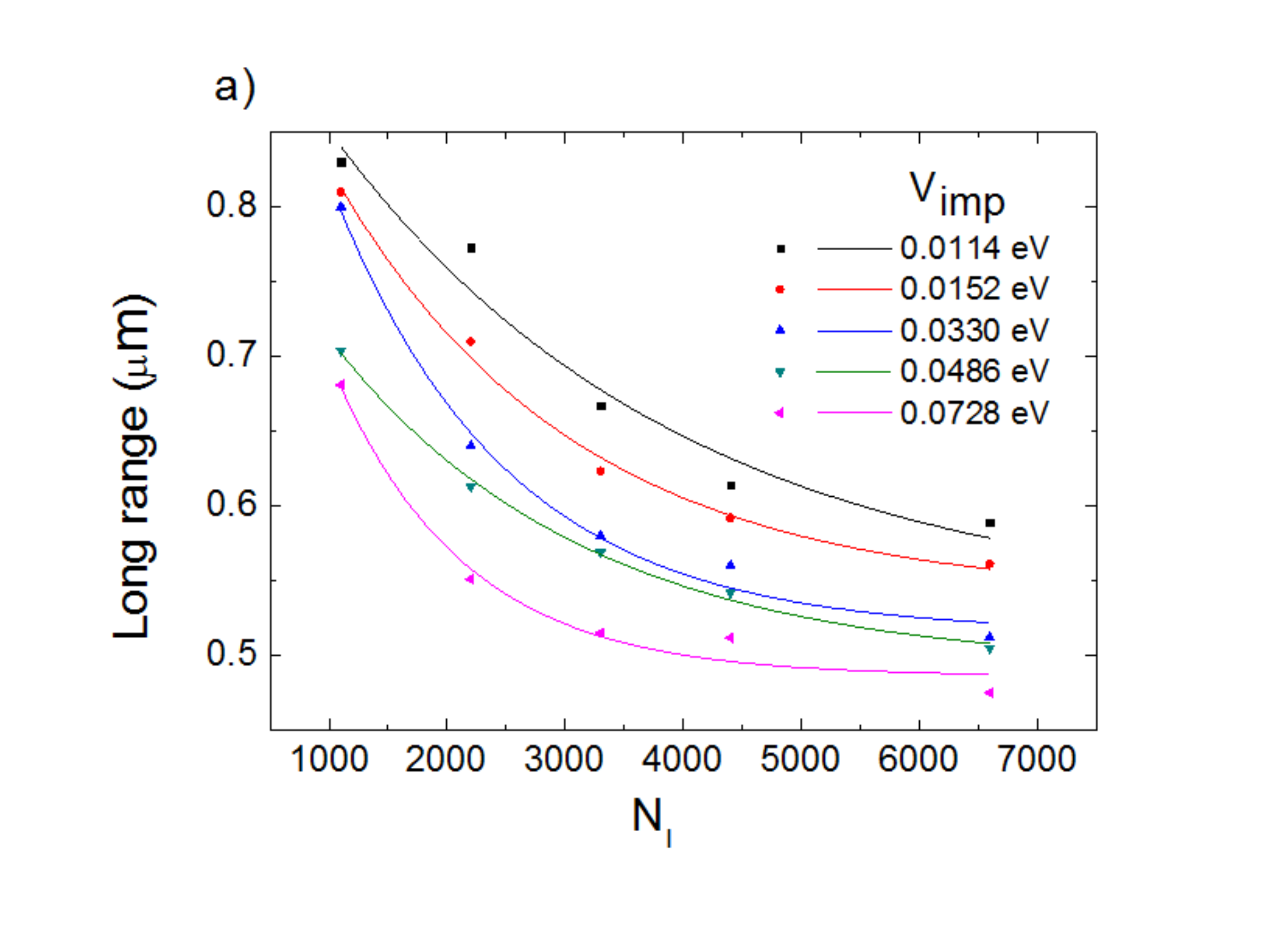}
\includegraphics[width=.5\linewidth,angle=0]{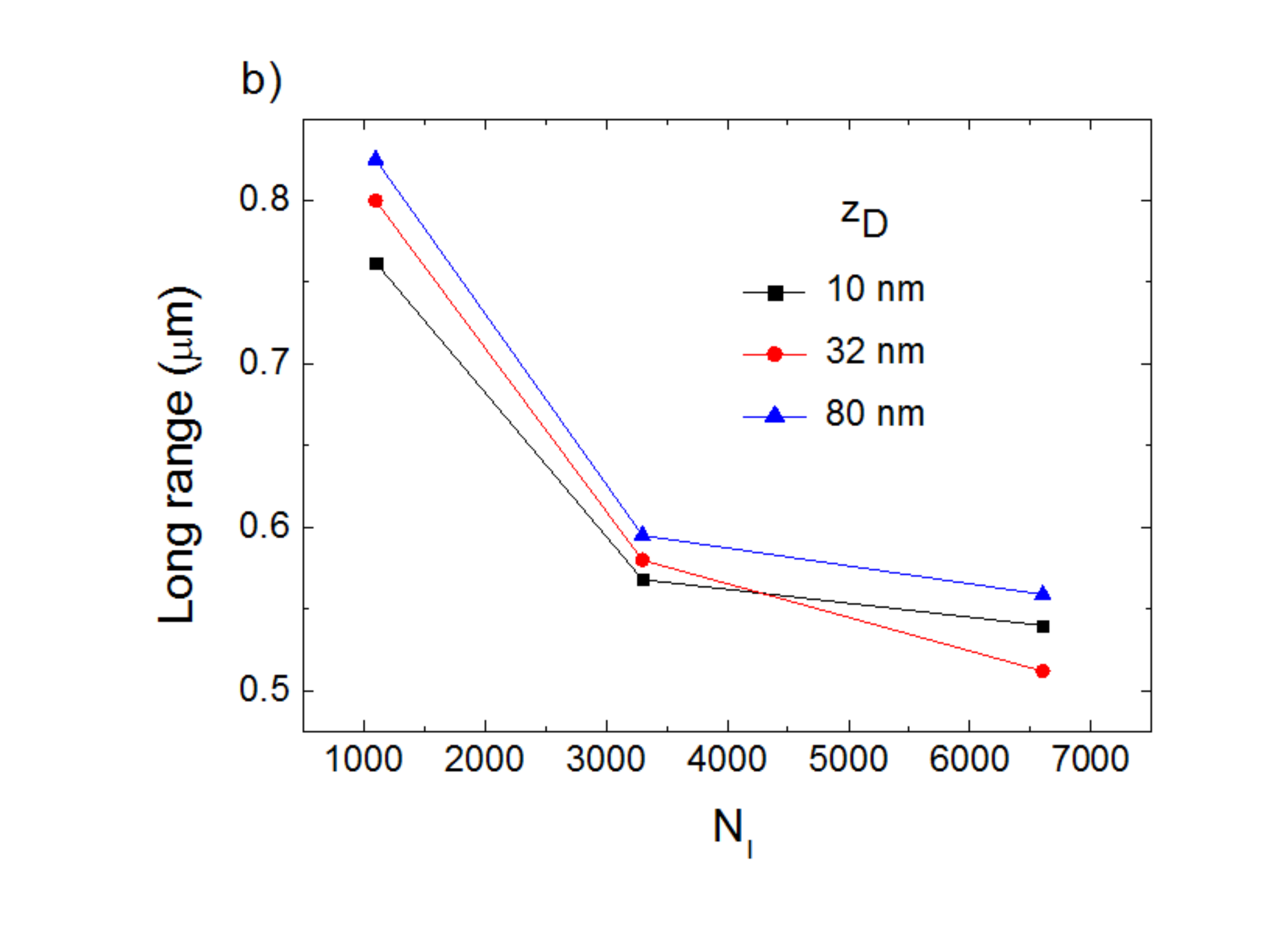}
\end{minipage}
\caption{Statistically estimated range of the density fluctuations
as a function of number of impurities, considering various
impurity strengths (a) and spacer thicknesses (b). The
calculations are done at zero temperature considering Coulomb
impurities. The long range potential fluctuations become larger
than the size of the unit cell if one considers less than $\%$5
disorder. \label{fig:8}}
\end{figure}

Next section is devoted to investigate the widths of the quantized
Hall plateaus utilizing our findings. We consider mainly two ``mobility'' regimes, where
the long range fluctuations is at the order of microns (high
mobility) and is at the order of few hundred nanometers, low
mobility. However, the amplitude of the total potential
fluctuations will be estimated not only depending on the number of
impurities but also depending on the spacer thickness, range and
amplitude of single impurity potential.
\section{Quantized Hall plateaus\label{quanHP}}
The main aim of this section is to provide a systematic
investigation of the quantized Hall plateau (QHP) widths within
the screening theory of the IQHE~\cite{siddiki2004}, therefore
here we summarize the essential findings of the mentioned theory.
In calculating the QHPs one needs to know local conductivities,
namely the longitudinal $\sigma_{\rm l}(x,y)$ and the transverse
$\sigma_{\rm H}(x,y)$. To determine these quantities it is
required to relate the electron density distribution $n_{\rm
el}(x,y)$ to the local conductivities explicitly. Here we utilize
the SCBA~\cite{Ando82:437}. However, the calculation of the
electron density and the potential distribution including direct
Coulomb interaction is not straightforward, one has to solve the
Schr\"odinger and the Poisson equations simultaneously. This is
done within the Thomas-Fermi approximation which provides the
following prescription to calculate the electron density \begin{equation}n_{\rm el}(x,y)=\int dE D(E)\frac{1}{e^{(E_F-V(x,y))/k_BT}+1},
\label{density}\end{equation}where $D(E)$ is the appropriate density of
states calculated within the SCBA, where $k_B$ is the Boltzmann
constant and $T$ temperature. The total potential is obtained from
\begin{equation}V(x,y)=\frac{2e^2}{\bar{\kappa}}\int dxdy K(x,y,x',y')n_{\rm
el}(x,y),\label{potential}\end{equation}and the Kernel $K(x,y,x',y')$ is the
solution of the Poisson equation satisfying the boundary
conditions to be discussed next.

In the following we assume a translation in variance in
$y$-direction and implement the boundary conditions $V(-d)=V(d)=0$
($2d$ being the sample width), proposed by Chklovskii
\emph{et.al.}~\cite{Chklovskii92:4026}, such a geometry allows us
to calculate the Kernel in a closed form. Hence,
Eqs.~(\ref{density}) and~(\ref{potential}) forms the self-consistency. For a given initial potential distribution, the electron
concentration can be calculated at finite temperature and magnetic
field, where the density of states $D(E)$ contains the information
about the quantizing magnetic field and the effect of short range
impurities. Here we implicitly assume that the electrons reside in
the interval $-b<x<b$ (where, $d_l=|d-b|/d$ is called the
depletion length), and is fixed by the Fermi energy, \emph{i.e.}
the number of electrons, hence donors. As a direct consequence of Landau
quantization and the locally varying electrostatic potential, the
electronic system is separated into two distinct regions, when
solving the above self-consistent equations iteratively: i) The
Fermi energy equals to (spin degenerate) Landau energy and due
to DOS the system illustrates a metallic behavior, the
compressible region, ii) The insulator like incompressible region,
where $E_F$ falls in between two consequent eigen-energies and no
states are available~\cite{Chklovskii92:4026,Siddiki03:125315}. It
is usual to define the filling factor $\nu$, to express the
electron density in terms of the applied $B$ field as, $\nu=2\pi
l^2 n_{\rm el}$. Since all the states below the Fermi energy are
occupied the filling factor of the incompressible regions
correspond to integer values (\emph{e.g.} $\nu=2,4,6...$), whereas
the compressible regions have non-integer values, due to partially
occupied higher most Landau level. The spatial distribution and
widths of these regions are determined by the confinement
potential~\cite{Chklovskii92:4026}, magnetic
field~\cite{Lier94:7757}, temperature~\cite{Oh97:13519} and
level broadening~\cite{Guven03:115327,siddiki2004}. For the purpose of the
present work we fix the confinement potential profile by
confining ourselves to the Chklovskii geometry and keeping the
donor concentration (and distribution) constant. Moreover we
perform our calculations at a default temperature given by
$k_BT/E_F^0=0.02$, where $E_F^0$ is the Fermi energy calculated
for the electron concentration at the center of the sample and is typically similar to 10 meV.

The next step is to calculate the global resistances, \emph{i.e.}
the longitudinal $R_{\rm L}$ and Hall $R_{\rm H}$ resistances,
starting from the local conductivity tensor elements. Such a
calculation is done within a relaxed local model that relates the
current densities $\textbf{j}(x,y)$ to the electric fields
$\textbf{E}(x,y)$, namely the local Ohm's law: \begin{equation}\textbf{j}(x,y)=\hat{\sigma}(x,y)\textbf{E}(x,y). \end{equation}The strict
locality of the conductivity model is lifted by an spatial
averaging process~\cite{siddiki2004} over the quantum mechanical
length scales and an averaged conductivity tensor
$\hat{\overline{\sigma}}(x,y)$ is used to obtain the global
resistances. It should be emphasized that, such an averaging
process also simulates the quantum mechanical effects on the
electrostatic quantities. To be explicit: if the widths of the
current carrying incompressible strips become narrower than the
extend of the wave functions, these strips become ``leaky'' which
can not decouple the two sides of the Hall bar and back-scattering
takes place. Therefore, to simulate the ``leakiness" of the
incompressible strips we perform coarse-graining over quantum
mechanical length scales.

\begin{figure}[!t]\center{

\includegraphics[width=.9\linewidth,angle=0]{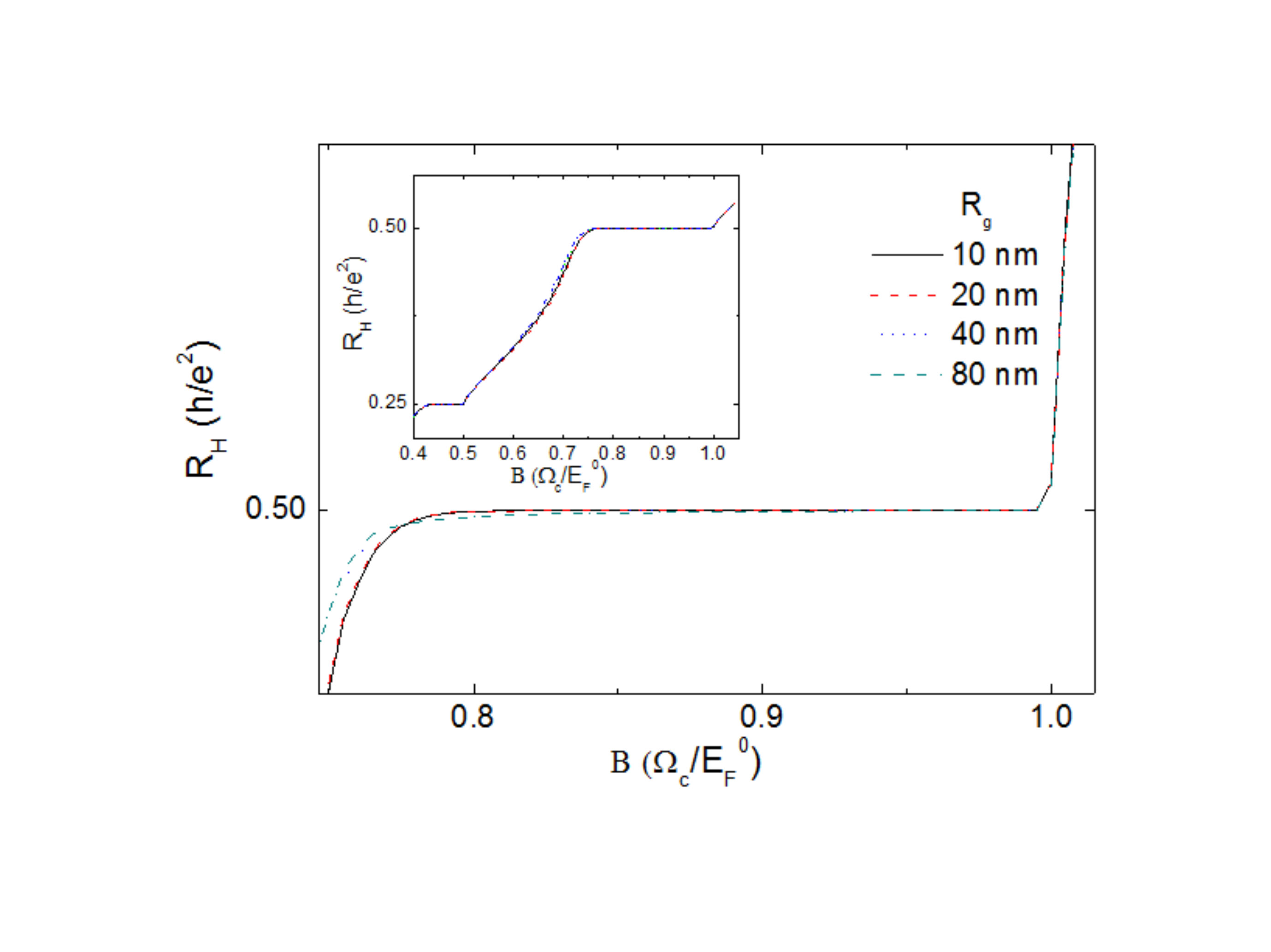}

\caption{The Hall resistances versus magnetic field, calculated at
default temperature and considering a 10$~\mu$m sample for
different ranges of the single impurity potential. Inset depicts a
larger $B$ field interval, where $\nu=4$ plateau can also be
observed. \label{fig:9}}}
\end{figure}

Now let us relate the local conductivities with the local filling
factors. Since the compressible regions behave like a metal within
these regions there is finite scattering leading to finite
conductivity. In contrast, within the incompressible regions the
back-scattering is absent, hence, the longitudinal conductivity
(and simultaneously resistivity) vanishes. Therefore, all the
imposed current is confined to these regions. The Hall
conductivity, meanwhile is just proportional to the local electron
density. The explicit forms of the conductivity tensor elements
are presented elsewhere~\cite{siddiki2004}. Having the electron
density and local magneto-transport coefficients at hand, we
perform calculations to obtain the widths of the quantized Hall
plateaus utilizing the above described, microscopic model assisted
by the local Ohm's law at a fixed external current $I$. Further
details of the calculation scheme is reviewed in
Ref.~\cite{Gerhardts08:378}.

\subsection{Single impurity potentials: Level broadening and conductivities \label{singleimp}}
Since the very early days of the charge transport theory,
collisions played an important role. Such a scattering based definition of conduction also
applies for the system at hand, \emph{i.e.} a two-dimensional
electron gas subject to perpendicular magnetic field. Among many
other approaches~\cite{Gerhardts75:285,Guven03:115327,TobiasK06:h} the SCBA emerged as a
reasonable model to describe the DOS assuming Gaussian impurities,
considering short range scattering. A single impurity has two distinct
parameters that represents the properties of the resulting
potential, the range $R_{\rm g}$ (at the order of separation
thickness) and the amplitude of the potential (in relevant units),
$\widetilde{V}_{\rm imp}$. However, these two parameters are not
enough to define the widths of the Landau levels ($\Gamma$), another
important parameter is the number of the impurities, $N_I$. In the
previous section we have already investigated these three
parameters in scope of potential landscape, now we utilize our
findings to define the level widths and the conductivities. It is
more convenient to write the single impurity potential of the
form,
\begin{table}\center{
\begin{tabular}{|c|c|c|c|c|}
\hline
$d_l=$ 70 nm  & $R_{\rm g}=$ 10 nm & 20 nm& 40 nm& 80 nm\\
\hline
\hline 2d= $2~\mu$m &0.120 & 0.120 & 0.100 & 0.050 \\
\hline $3~\mu$m& 0.135 & 0.125 & 0.090 & 0.035 \\
\hline $5~\mu$m&0.140 & 0.115 & 0.070 & 0.020 \\
\hline $8~\mu$m& 0.135 & 0.095 & 0.050 & 0.010 \\
\hline $10~\mu$m& 0.130 & 0.085 & 0.040 & 0.010 \\ \hline
\end{tabular}}
\end{table}
\begin{table}
\center{
\begin{tabular}{|c|c|c|c|c|}
\hline
$d_l=$ 150 nm & $R_{\rm g}=$ 10 nm & 20 nm& 40 nm& 80 nm\\
\hline
\hline 2d= $2~\mu$m &0.140 & 0.140 & 0.125 & 0.075 \\
\hline $3~\mu$m& 0.160 & 0.150 & 0.120 & 0.055 \\  \hline$5~\mu$m&
0.180 &
0.150 & 0.095 & 0.035 \\  \hline $8~\mu$m& 0.180 & 0.130 & 0.070 & 0.020 \\
\hline $10~\mu$m& 0.175 & 0.120 & 0.060 & 0.015 \\ \hline
\end{tabular}
\caption{The $\nu=2$ plateau widths obtained at default
temperature for two depletion lengths $d_l$ (left 75 nm, right 150
nm), while $\gamma_I=0.05$ is fixed (defined in
Eq.~\ref{eq:gammaI} and the related text below). The widths are
given in units of $\hbar\omega_c/E_F^0=\Omega_c/E_F^0$.}
\label{table:1}}
\end{table}
\begin{equation}V_{\rm g}(r)=\frac{\widetilde{V}_{\rm imp}}{\pi R_{\rm
g}^2}\exp{(-\frac{r^2}{R_{\rm g}^2})}. \end{equation}Together with the
impurity concentration, the relaxation time is defined as $\tau_0=\frac{\hbar^3}{N_I\widetilde{V}_{\rm imp}^2m^*}$ and in the
limit of delta impurities (\emph{i.e.} $R_{\rm g}\rightarrow0$)
the Landau level width $\Gamma$ takes the form $\Gamma=\sqrt{\frac{4N_I\widetilde{V}_{\rm imp}^2}{2 \pi l^2}}$. It is useful to
define the impurity strength parameter to investigate the effect of disorder by \begin{equation}\gamma_I^2=(\Gamma/\hbar \omega_c)^2= \frac{2N_I\widetilde{V}_{\rm
imp}^2m}{\pi \hbar^3 \omega_c}, \label{eq:gammaI}\end{equation} given in units of magnetic
energy $\hbar \omega_c=\frac{\hbar eB}{m}=\Omega_c$ and as a normalization parameter we
fix the magnetic energy at 10 T.

At this point we would like to make a remark on the concepts
short/long range impurities and short/long range potential
fluctuations, which is commonly mixed. By short range impurity
potential we mean that $R_{\rm g}\lesssim l$, however, by short
range potential fluctuation a length scale of the order of
$200-300$ nm is meant. The long range impurity potential
corresponds to $R_{\rm g}> l$ and long range potential
fluctuation is of the order of micrometers. Thus, when considering
short range impurities the potential fluctuations may be long
range, if $N_I$ is not large ($< \% 5$ of the donor
concentration). We have also observed that, the long-range
potential fluctuations are more efficiently screened by the 2DES
and their range can be at the order of 500 nm at most, when
assuming large impurity concentration, \emph{i.e.} $N_I>\% 10$.
\begin{figure}[!t]
\centering{
\includegraphics[width=.9\linewidth,angle=0]{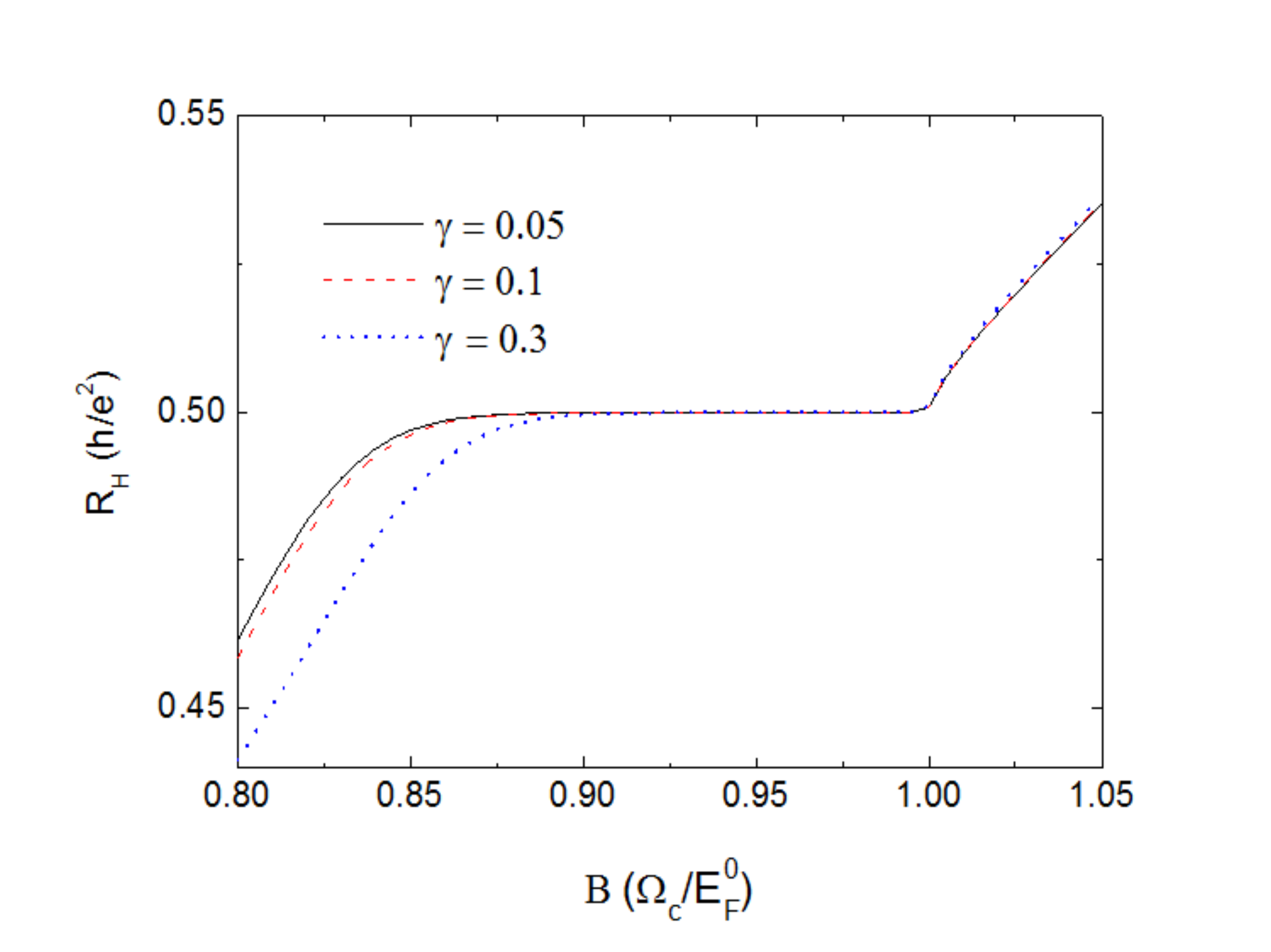}

\caption{The calculated Hall resistances at default temperature
assuming a 5 $\mu$m sample considering three characteristic value
of broadening parameter. The lowest mobility ($\gamma_I$=0.3)
shows the narrowest plateau. \label{fig:10}}}
\end{figure}
In light of the above findings and formulation we now investigate
the widths of the quantized Hall plateaus. Figure ~\ref{fig:9}
presents the calculated Hall resistances at a fixed temperature
for typical single impurity ranges. We observe that, when
increasing $R_{\rm g}$ the plateau widths remain approximately the
same, with a small variation, which is in contrast to the
experimental findings, \emph{i.e.} if the system is low mobility
(small $R_{\rm g}\Rightarrow$ highly broadened DOS) the plateau
are larger. In fact changing $R_{\rm g}$ from 10 nm to 20 nm
should increase the zero $B$ field mobility almost an order of
magnitude, when fixing the other parameters (see \emph{e.g} table I of
Ref.~\cite{siddiki2004}). The contradicting behavior is due
to the fact that the levels become broader when increasing the
single impurity range, therefore the incompressible strips become
narrower, which results in a narrower plateau. However, the long
range potential fluctuations are completely neglected, therefore
the effect(s) of disorder on the quantized Hall plateaus cannot be
described in a complete manner. To investigate the effect of the
single impurity range we systematically calculated the plateau
widths; table~\ref{table:1} depicts the calculated widths of the
Hall plateaus considering different sample widths, depletion
lengths, filling factors and $R_{\rm g}$. One sees that the
plateau widths are affected by the increase of impurity range,
however, in a completely wrong direction,\emph{ i.e.} plateaus
become narrower when decreasing the mobility. As we show in the
next section, it is not sufficient to describe mobility only
considering the range of a single impurity. Moreover, we also show
that the other two parameters defining $B=0$ mobility are either
not important or behaves in the opposite direction when
calculating the resistances.

Next we investigate the effect of the remaining two parameters,
$\widetilde{V}_{\rm imp}$ and $N_I$. However, these two parameters
both effect the level width simultaneously, thereby the widths of
the incompressible strips. Hence, one cannot to distinguish their
influence on the QHPs separately. Typical Hall resistances are
shown in Fig.~\ref{fig:10} calculated at default temperature
considering different impurity parameters. Similar to the range
parameter, we observe that the plateau widths become narrower when
the mobility is low, which also points that our single particle
based level broadening calculations are not in the correct
direction. Such a behavior is easy to understand, when we decrease
the mobility either by increasing the impurity concentration or by
the amplitude of the impurity potential, the Landau levels become
broader due to collisions. This means that, both the energetic and
spatial gap between two consequent levels is reduced, hence the
resulting incompressible strips are also narrower and fragile even
at low temperatures. A detailed investigation on the
incompressible widths depending on impurity parameters are
reported in Ref.~\cite{Guven03:115327}. It is known that if
there exists an incompressible strip wider than the Fermi
wavelength the system is in the quantized Hall
regime~\cite{siddiki2004}, therefore, if the gap is reduced the
incompressible strips are smeared, thus the quantized Hall plateau
vanish. As a general remark on the single particle theories, we
should note that such a reduced gap is also a gross problem for
the non-interacting
models~\cite{Laughlin81,Buettiker88:317,Halperin82:2185}, however,
one can overcome this discrepancy by making localization
assumptions~\cite{Kramer03:172}. Namely, one assumes that even
within the broadened Landau levels there are states, which are
localized, therefore electrons cannot contribute to transport.
Hence, although the gap is small (levels are broad) these
localized states serves as a reasonable candidate to explain the
low mobility behavior. In the early days of IQHE it was a great
challenge to describe and observe these localized
states~\cite{Cai86:3967}. Recent
experiments~\cite{Ahlswede01:562,Yacoby00:3133,josePHYSE,Ashoori05:136804}
show clearly that, the localization assumptions are not relevant
in all the cases, \emph{i.e.} narrow and high mobility samples.
Moreover, the universal behavior of the localization length
dictated by these theories fail~\cite{criticalexpo:09}. An
explicit treatment of the activation energy~\cite{Serpilactivation:09} and critical exponents are left to an
other publication. \subsection{Size effects on plateau widths}
\begin{figure}[!t]\centering
\includegraphics[width=.8\linewidth,angle=0]{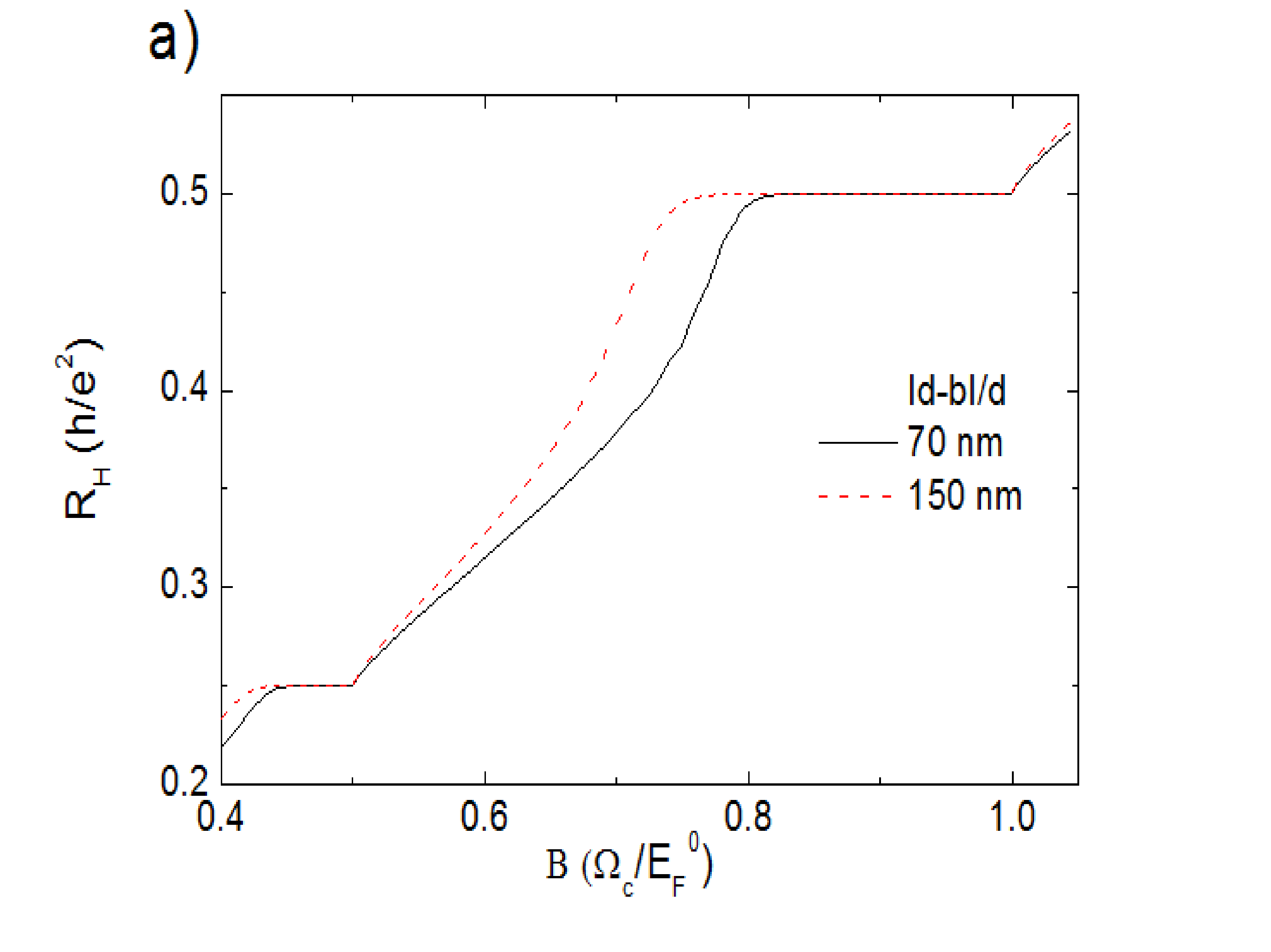}
\includegraphics[width=.8\linewidth,angle=0]{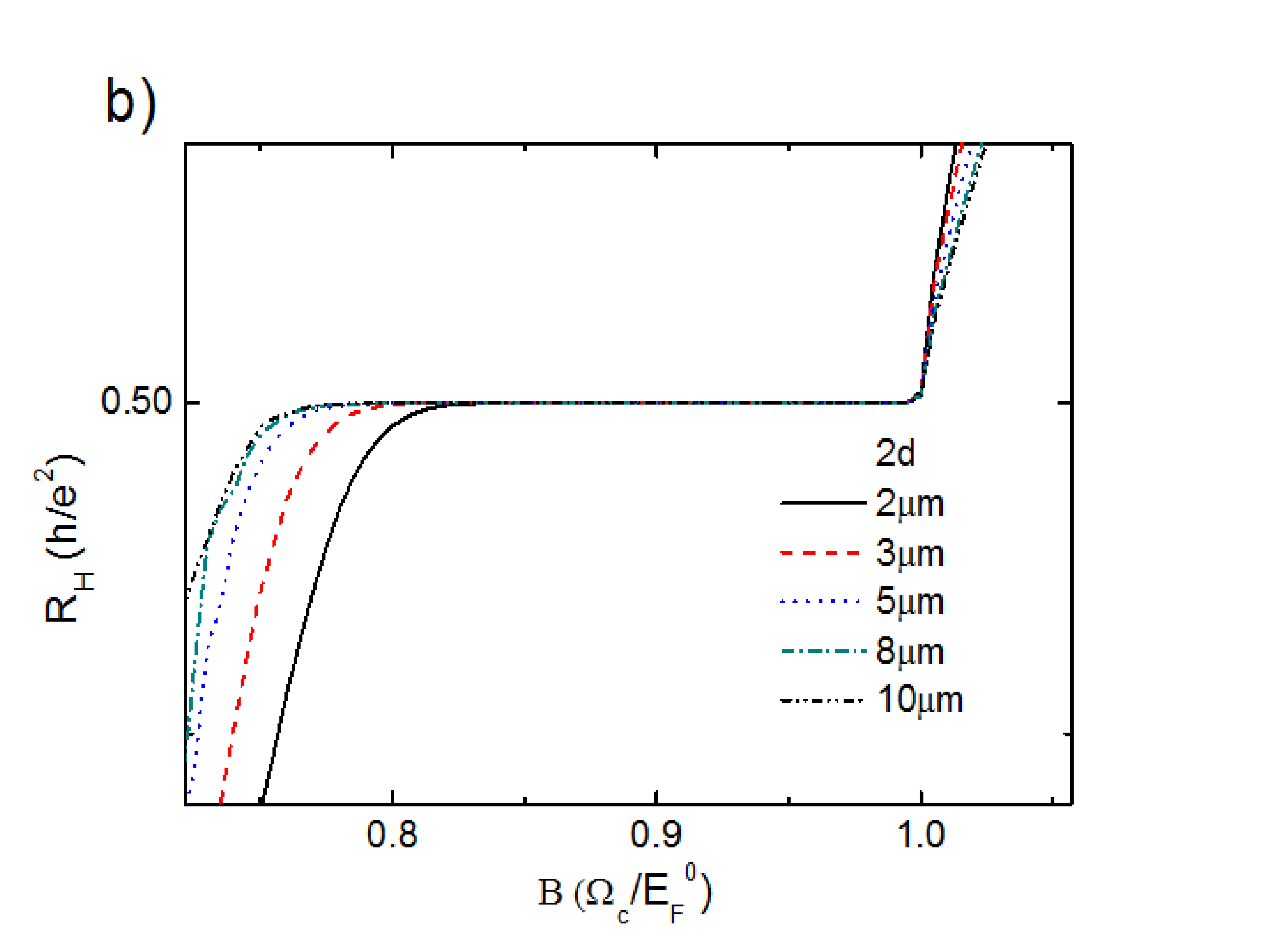}

\caption{a) The calculated Hall resistances at a large $B$ interval
at default temperature, setting $2d=5~\mu$m, $R_{\rm g}=20$ nm and
$\gamma_I=0.05$, while changing the depletion length. It is
clearly seen that depletion length is much more important than the
single impurity parameters in determining the plateau widths. (b) The direct comparison of the plateau widths considering
different sample sizes. The impurity parameters and depletion
lengths are kept constant. Calculations are done at
$k_BT/E_F^0=0.02$, whereas the donor density is $4\times10^{15}$
m$^{-2}$ for all sample sizes.
\label{fig:11}}
\end{figure}
Another important parameter in defining the plateau widths is the
depletion length $d_l$. The slope of the confinement potential
close to the edges essentially determines the widths of the
incompressible strips~\cite{Chklovskii92:4026}, which in turn determines the plateau
widths. In Fig.~\ref{fig:11} we show the $\nu$=2 plateau calculated
for two different depletion lengths, we see that for the larger
depletion the plateau is more extended. Since, the larger the
depletion is, the smoother the electron density is. Therefore,
resulting incompressible strips are wider, hence the plateau. Such an argument will fail if one considers a
highly disordered large sample, which we discuss in
Sec.~\ref{manyimp}. Next, we compare the plateau widths of
different sample sizes while keeping constant the disorder
parameters and depletion length. Figure ~\ref{fig:11} depicts the
sample size dependency of $\nu=2$ plateau width. It is seen that
the larger samples present wider plateaus, if the magnetic field
is normalized with the center Fermi energy, $E_F^0$. One can
understand this by similar arguments given above, \emph{i.e.} if
the sample is narrow the variation of the confinement potential is
stronger, therefore the incompressible strips become narrower,
hence, the plateaus. The discrepancy between the experimental
results and the screening theory of the IQHE is solved if one
considers not only the single impurity potentials but also the
overall \emph{disorder} potential landscape generated by the
impurities. In the next part of this section, we investigate the
effect of the long range potential fluctuations on the quantized
Hall plateaus and find that, when the mobility is reduced the
plateaus become wider and stabile, as it is observed in many
experiments, (see \emph{e.g.}
Refs.~\cite{Haug87:SdH,josePHYSE,jose:prl}).
\begin{figure}[!t]
\begin{minipage}{1\linewidth}\centering
\includegraphics[width=.70\linewidth,angle=0]{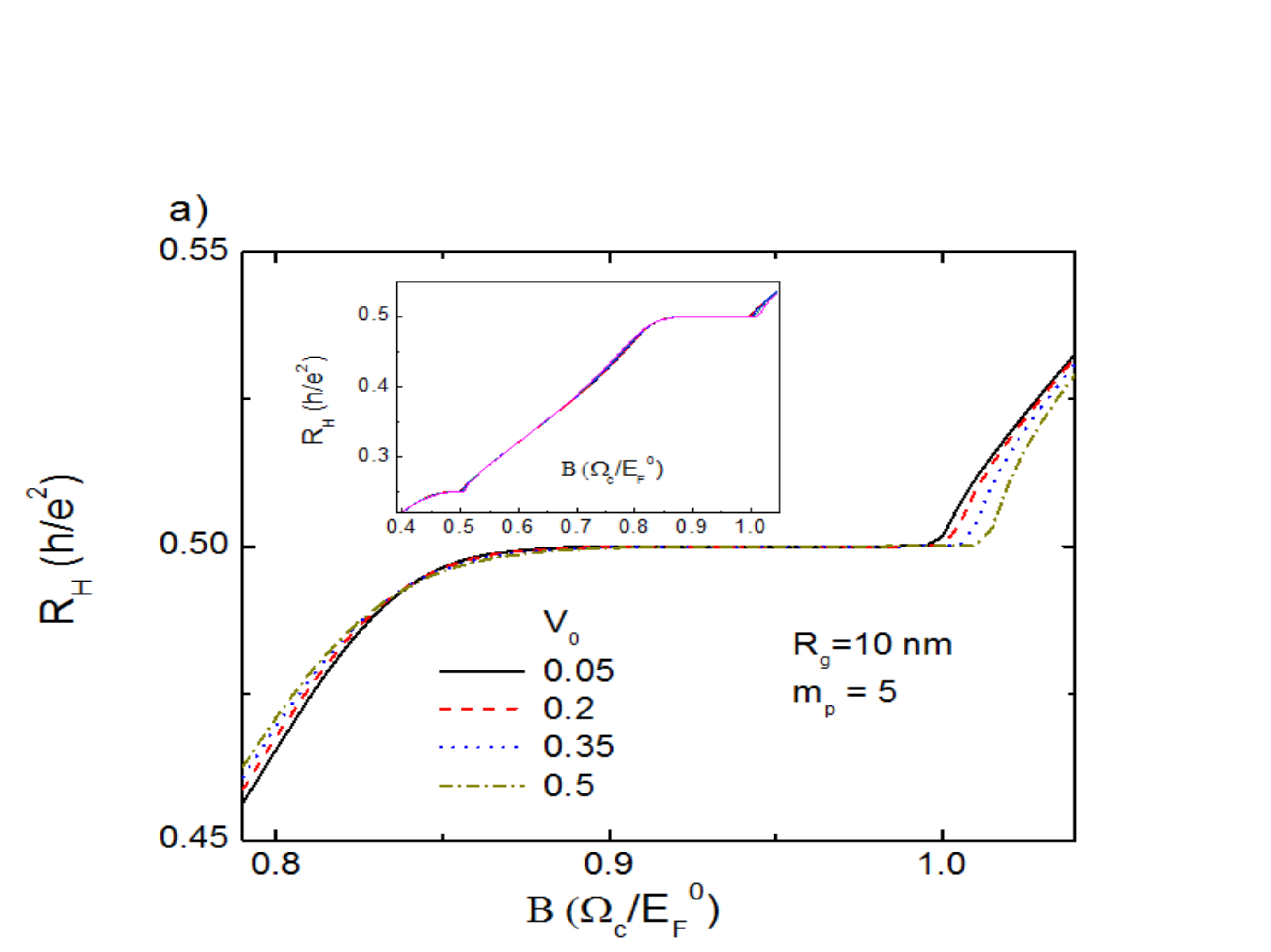}
\end{minipage}

\caption{Self-consistently obtained Hall resistances for a
modulated system considering a sample of 3 $\mu$m. The depletion
lengths and other single impurity parameters are kept fixed,
whereas the parity of the modulation period is set 5.\label{fig:13}}

\end{figure}
\subsection{Many many impurities: Potential fluctuations\label{manyimp}}
So far we have investigated the effect of single impurity
potentials on the overall potential landscape in Sec.~\ref{treed}
and on the widths of the plateaus in Sec.~\ref{singleimp}. We have
seen that, at high impurity concentration the overall potential
fluctuates over a length scale of couple of hundred nanometers,
whereas for low $N_I$ concentration such length scale can be as
large as micrometers. Now we include the effect of this long range
potential fluctuations \emph{into} our screening calculations via
modulation potential defined as $V_{\rm mod}(x)=V_0\cos{(\frac{
2 \pi x m_{p}}{2d}})$ where, the modulation period $m_{p}$, is
chosen such that the boundary conditions are preserved. At the
moment, we consider two modulation periods regardless of the
sample width and vary the amplitude of the modulation potential.
In the next section, however, we select these two parameters
from our estimations obtained in Sec.~\ref{impurity} and
Sec.~\ref{treed}.

Figure~\ref{fig:13} depicts the self-consistently calculated Hall
resistances, considering different modulation amplitudes $V_0$ for
a fixed sample width ($2d=3$ $\mu$m) and $m_{p}=5$. We observe
that, the plateaus become wider from the high $B$ field side, when
$V_0$ is increased, \emph{i.e} mobility is reduced. Such a
behavior is now consistent with the experimental findings. Since
the QHPs occur whenever an incompressible strip is formed
(somewhere) in the sample and the modulation forces the 2DES to
form an incompressible strip at a higher magnetic field, therefore
the plateau is also extended up to higher field compared with the
(approximately) non-modulated calculation, $V_0/E_F^0<0.1$.

Our investigation of the impurities lead us to conclude
that, one has to define mobility at high magnetic fields also
taking into account screening effects in general and furthermore
also the geometric properties of the sample such as the width and
depletion length. As an example if we consider an impurity
concentration of $\approx\% 1$ the long range part of the
potential fluctuation can be approximated to 900 nm. However, note
that the amplitude of this fluctuation varies between $\% 5-25$ of
the Fermi energy, considering different separation thicknesses, therefore the wafer
changes from low mobility to intermediate one. Another important
parameter is the number of modulations within the system: a sample
with an extend of $2~\mu$m and $V_{0}/E_F=0.1$ is a high mobility
sample with the same $m_p$ (only 2 maximum), however, sample with
a width of $10~\mu$m is low mobility (10 maximum). In the next
section we study the plateau widths of different mobility samples,
while keeping constant the extend and the amplitude of long range
potential fluctuations (\emph{i.e.} $V_0$ and $m_p$) and short
range impurity parameters ($\widetilde{V}_{\rm imp}$, $N_I$ and
$R_{\rm g}$) under experimental conditions.
\section{Discussion:Comparison with the experiments\label{exper}}
In this final section, we harvest our findings of the previous
sections to make quantitative estimations of the plateau widths,
considering narrow gate defined samples. Our aim is to show the
qualitative and quantitative differences between ``high'' and
``low'' mobility samples, by taking into account properties of the
single impurity potentials and the resulting disorder potential.
The experimental realizations of these samples are reported in the
literature~\cite{josePHYSE,jose:prl}. We estimated in
Sec.~\ref{treed} that, the range of the potential fluctuations is
$\lesssim 500$ nm for low mobility ($N_I>3300$) and is $\gtrsim
1~\mu$m at high mobility. Therefore, the modulation period is
chosen such that many oscillations correspond to low mobility, and
few oscillations correspond high mobility. As an specific example
let us consider a $10~\mu$m sample, for the low mobility we choose
$m_p=19-20$ and for the high mobility $m_p$ is taken as 9 or 10.
The amplitude of the disorder potential is damped to $\% 50$ of
the Fermi energy when considering the effect of spacer thickness,
however, including screening this amplitude is further reduced to
few percents. In light of this estimations the low mobility will
be presented by a modulation amplitude of $V_0/E_F^0=0.5$, whereas
high mobility corresponds to $V_0/E_F^0=0.05$. Therefore, we have
4 different combinations of the disorder potential parameters
yielding four different mobilities considering two sample widths,
as tabulated in table ~\ref{table:2}. The second important aspect
of the disorder is the single impurity parameters, for low
mobility set we choose $R_{\rm g}=20$ nm and $\gamma_I=0.3$,
whereas for high mobility $R_{\rm g}=10$ nm and $\gamma_I=0.05$ is
set. Remember that, the range of the single impurity is much less
important than $\gamma_I$ in determining the plateau width (see
sec.~\ref{singleimp}).
\begin{table}

\begin{tabular}{|c|c|c|c|}
  \hline
  mobility & $m_p$ (10 $\mu$m) & $m_p$ (2 $\mu$m) & $V_0/E_F^0$ \\
  \hline
  \hline
  low & 19-20& 5-6 & 0.5 \\
  \hline
  intermediate 1 & 9-10 & 2-3 & 0.5 \\
  \hline
  intermediate 2 & 19-20 & 5-6& 0.05 \\
  \hline
  high & 9-10 & 2-3 & 0.05 \\
  \hline
\end{tabular}
\centering \caption{A qualitative comparison of the mobility in
the presence of magnetic field also taking into account
self-consistent screening. Mobility also depends on the size of
the sample when screening is also considered.}\label{table:2}
\end{table}

\begin{figure}[!t]
\centering{
\includegraphics[width=.9\linewidth,angle=0]{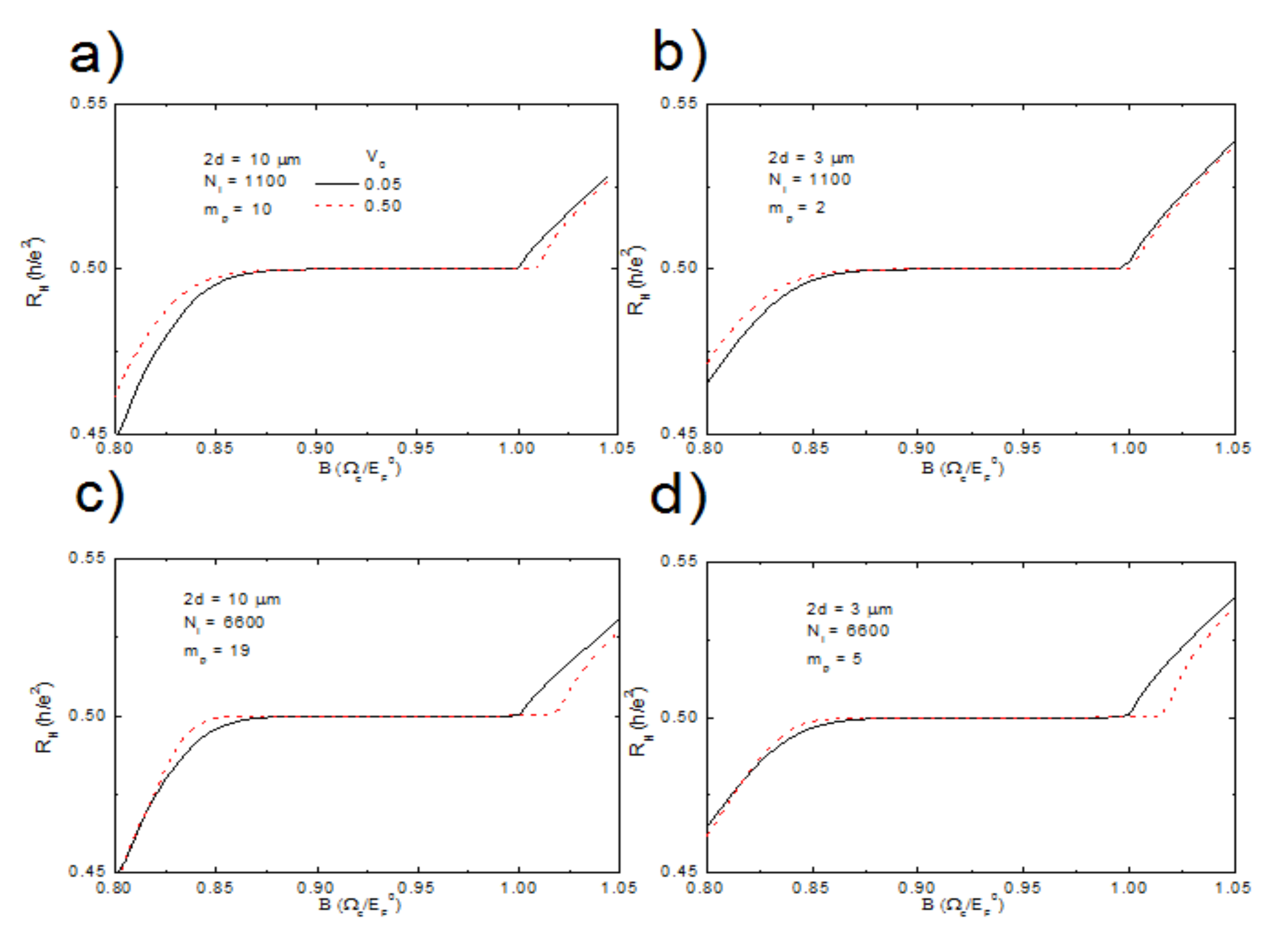}

\caption{Line plots of the Hall resistance as a function of
magnetic field considering two sample widths ($2d=10~\mu$m left
panels, $2d=3~\mu$m right panels) and impurity concentrations
($\sim\%3$ (a) and (b), $\sim \%20$ (c) and (d)). Here the single
impurity parameters are calculated from Eq.~\ref{eq:gammaI},
otherwise other parameters are the same. \label{fig:15}}}
\end{figure}

Figure~\ref{fig:15} summarizes our results considering above
discussed mobility regimes for two different sample widths. In
Fig.~\ref{fig:15}a, we show the calculated Hall resistances for a
sample of 10 microns with the highest mobility (solid (black)
line) and intermediate 1 mobility (broken (red) line). The solid
line is the highest mobility since the range of the fluctuations
are at the order of 1 $\mu$m and the amplitude of the modulation
potential is five percent of the Fermi energy. The broken line
presents the intermediate mobility considering a modulation
amplitude of fifty percent. We observe that the lower mobility
wafer presents a larger quantized Hall plateau, which is now in
complete agreement with the experimental results. Moreover, our
calculation scheme is free of localization assumptions in contrast
to the known literature and we only considered a very limited
level broadening, \emph{i.e.} $\gamma_I=0.05$. In fact our results
also hold for Dirac-delta Landau levels, however, for the sake of
consistency we choose the broadening parameters according to the
selected disorder parameters. In Fig.~\ref{fig:15}c, we show two
curves for even lower mobilities, the solid line corresponds to
the intermediate 2 case, whereas the broken line is the lowest
mobility considered here. The potential fluctuation range
(\emph{i.e.} the modulation period) is chosen to present the low
mobility wafer. We again see that for the lowest mobility the
quantized Hall plateau is enlarged considerably from both edges of
the plateau. These results explicitly show that the quantized Hall
plateaus become broader if one strongly modulates the electronic
system by long range potential fluctuations, either by changing
the range \emph{or} the amplitude of the modulation. Similar
results are also obtained for a relatively narrower sample
$2d=3~\mu$m, Fig.~\ref{fig:15}b and ~\ref{fig:15}d, however, we
see that decreasing the range of the potential fluctuation is more
efficient in enlarging the quantized Hall plateaus when compared
to the effect of the amplitude of the modulation.

\begin{figure}[!t]
\centering{
\includegraphics[width=.5\linewidth,angle=0]{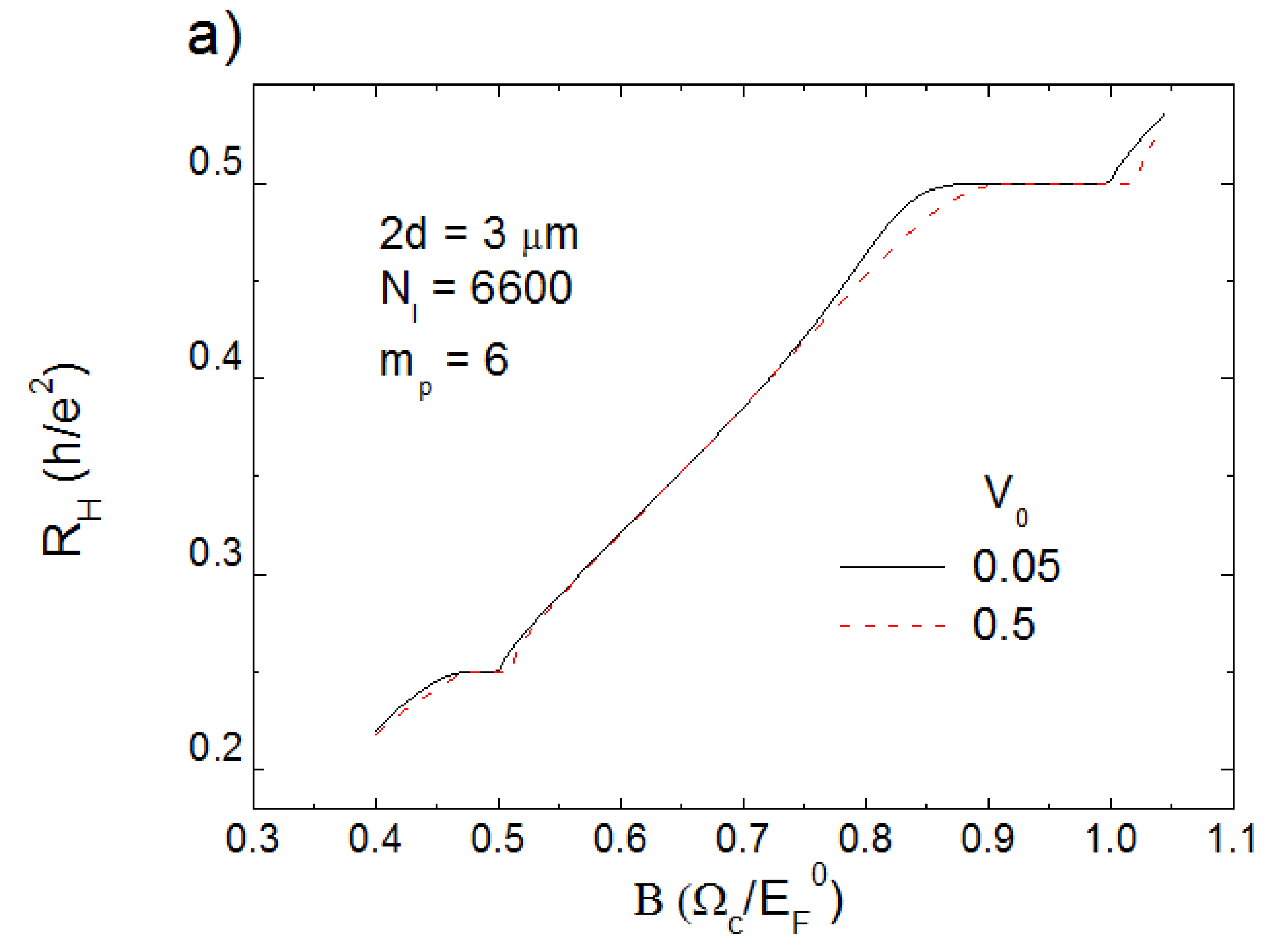}
\includegraphics[width=.5\linewidth,angle=0]{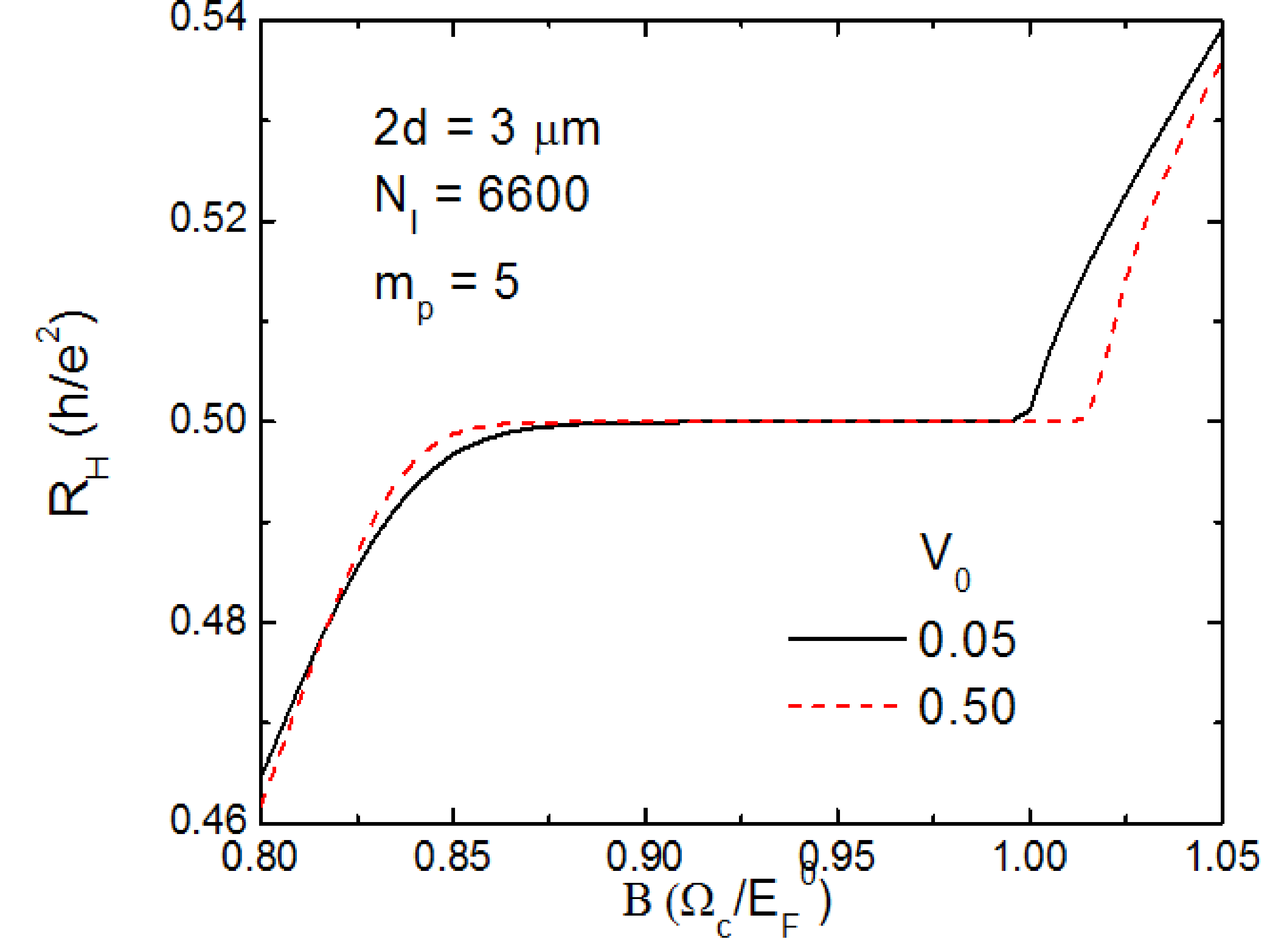}}

\caption{Even-odd parity dependency of the Hall plateaus at high
impurity concentration. (a) corresponds to a ``acceptor'' doped
wafer, whereas in (b) the ionized impurities are positively
charged. \label{fig:16}}
\end{figure}

The last interesting investigation is on the parity of the
modulation period, \emph{i.e.} whether $m_p$ is odd or even.
Figure~\ref{fig:16} presents the different behavior when
considering even (a) or odd (b) periods. Here, all the disorder
parameters are kept fixed, other than the parity. We see that for
the even parity the plateau is shifted towards the high field
edge, both for $\nu=2$ and 4, whereas for the odd parity the
plateau is enlarged from both sides. This tendency is also
observed for the larger sample (not shown here). We attribute this
behavior again to the formation of the incompressible strips,
however, this time only to the one residing at the center of the
sample, \emph{i.e.} the bulk incompressible strip. The picture is as follows: If the maxima of the modulation
potential is at the center of the sample, the incompressible strip
is formed at a higher magnetic field value, whereas, the edge
incompressible strips become narrower at the lower field side.
Hence, due to the larger incompressible strip at the bulk of the
sample the plateau is shifted to the higher field, in contrast,
due to the narrower (compared to the unmodulated system) edge
strips the plateau is cut off at higher fields. Since, the edge
incompressible strip becomes narrower than the extend of the wave
function. For the odd parity, the edge incompressible strips
become wider, therefore, the plateau extends to the lower $B$
fields. The enhancement at the high field edge results from the
two maximum in the proximity of the center. For a better
visualization of the incompressible strip distribution we suggest
reader to look at Fig.2 of Ref.~\cite{Siddiki:ijmp} and
Fig.1 of Ref.~\cite{Siddiki02:Oxford}. Such a shift of the
quantized Hall plateaus is also reported in the
literature~\cite{Haug87:SdH} and is attributed to the asymmetrical
density of states due to the acceptors in the
system~\cite{Gerhardts87:asymmetry}. We claim that, the shift due
to the modulation parity change observed in our calculations
overlap with their findings. Note that in our calculations we only
consider symmetric DOS, however, replacing a maxima with a minima
at the confinement potential corresponds to the acceptor behavior
of the dopants. A systematic experimental investigation is
suggested to understand the underlying physical mechanism, where
the system is doped with small number of acceptors.
\section{Conclusion}
In this work we tackled with the long standing and widely
discussed question of the effect of disorder on the quantized Hall plateaus. The distinguishing aspect of our approach relies on the
separate treatment of the long and short range of the disorder
potential. We show that assuming Gaussian impurities is not
sufficient to describe long range potential fluctuations, however,
is adequate to give a prescription in defining the density of
states broadening and conductivities. The discrepancy in handling
the long range potential fluctuations is cured by the inclusion of
a modulation potential to the self-consistent calculations. We
estimated the range of these fluctuations from our analytical and
numerical calculations considering the effect of dielectric spacer
and the screening of the 2DES. It is observed that spacer damps
the short range fluctuations effectively, whereas the direct
Coulomb interaction is dominant in screening the long range
fluctuations. Utilizing the estimations of the range and the
amplitude of potential fluctuations, we classified mobility in
four groups and calculated the Hall resistances within the
screening theory of the quantized Hall effect. We found that the
Hall plateaus are wider when decreasing the mobility, not
surprisingly. However, the most important point of our theory is
that, we do not consider any localization assumptions, still
obtain correct behavior of the plateau widths. We show that $B=0$
and/or short range impurity defined mobility is not adequate to
describe the actual mobility at high magnetic fields, moreover,
one has to include geometrical properties of the sample at hand.

A natural persecutor theoretical investigation of the present work
should deal with the activated behavior of the longitudinal
resistance within the screening theory. As it is well known, the
properties of the localized states, \emph{e.g}. the localization length,
is usually obtained from the activation experiments~\cite{Matthews05:497}. Moreover,
spin generalization of the screening theory~\cite{afifPHYSEspin} is necessary to
describe and investigate odd integer quantized plateaus also considering level broadening, namely disorder.
\ack
One of the authors (A.S.) would like to thank E. Ahlswede, S. C.
Lok and J. Weiss for the enlightening discussions on the disorder
from ``an experimentalist'' point of view. The Authors
acknowledges, the Feza-Grsey Institute for supporting the III.
Nano-electronic symposium, where this work has been conducted
partially and would like to acknowledge the Scientific and
Technical Research Council of Turkey (TUBITAK) for supporting
under grant no 109T083.

\section*{References}

\begin{thebibliography}{10}

\bibitem{Kramer03:172}
B.~Kramer, S.~Kettemann, and T.~Ohtsuki.
\newblock Localization in the quantum hall regime.
\newblock {\em Physica E}, 20:172, 2003.

\bibitem{Schweitzer85:379}
L.~Schweitzer, B.~Kramer, and A.~MacKinnon.
\newblock Selective probing of ballistic electron orbits in rectangular antidot
  lattices.
\newblock {\em Z. Physik B}, 59:379, 1985.

\bibitem{Cai86:3967}
W.~Cai and C.~S. Ting.
\newblock Screening effect on the landau-level broadening for electrons in
  gaas-gaalas heterostructures.
\newblock {\em Phys. Rev. B}, 33:3967, 1986.

\bibitem{Ando74:959}
T.~Ando and Y.~Uemura.
\newblock {\em J. Phys. Soc. Japan}, 36:959, 1974.

\bibitem{Ando75:279}
T.~Ando, Y.~Matsumoto, and Y.~Uemura.
\newblock {\em J. Phys. Soc. Japan}, 39:279, 1975.

\bibitem{Nixon:90}
J.~A. {Nixon} and J.~H. {Davies}.
\newblock {Potential fluctuations in heterostructure devices}.
\newblock {\em Phys. Rev. B}, 41:7929--7932, April 1990.

\bibitem{stopadisorder:96}
M.~{Stopa} and Y.~{Aoyagi}.
\newblock {Effect of donor layer ordering on the formation of single mode
  quantum wires}.
\newblock {\em Physica B Condensed Matter}, 227:61--64, September 1996.

\bibitem{Fogler94:1656}
M.~M. Fogler and B.~I. Shklovskii.
\newblock Resistance of a long wire in the quantum hall regime.
\newblock {\em Phys. Rev. B.}, 50:1656, 1994.

\bibitem{Efros88:1019}
A.~L. Efros.
\newblock Non-linear screening and the background density of 2deg states in
  magnetic field.
\newblock {\em Solid State Commun.}, 67:1019, 1988.

\bibitem{Buettiker86:1761}
M.~B{\"u}ttiker.
\newblock Four-terminal phase-coherent conductance.
\newblock {\em Phys. Rev. Lett.}, 57:1761, 1986.

\bibitem{Chklovskii92:4026}
D.~B. Chklovskii, B.~I. Shklovskii, and L.~I. Glazman.
\newblock Electrostatics of edge states.
\newblock {\em Phys. Rev. B}, 46:4026, 1992.

\bibitem{Datta}
S.~Datta.
\newblock In {\em Electronic Transport in Mesoscopic Systems}, Cambridge, 1995.
  University press.

\bibitem{Wilde06:disorder}
N.~{Ruhe}, J.~I. {Springborn}, C.~{Heyn}, M.~A. {Wilde}, and D.~{Grundler}.
\newblock {Simultaneous measurement of the de Haas-van Alphen and the
  Shubnikov-de Haas effect in a two-dimensional electron system}.
\newblock {\em Phys. Rev. B}, 74(23):235326--+, December 2006.

\bibitem{Mares:09}
Jiri~J Mares, Afif Siddiki, Dobroslav Kindl, Pavel Hubik, and Jozef Kristofik.
\newblock Electrostatic screening and experimental evidence of a topological
  phase transition in a bulk quantum hall liquid.
\newblock {\em New Journal of Physics}, 11(8):083028, 2009.

\bibitem{josePHYSE}
J.~{Horas}, A.~{Siddiki}, J.~{Moser}, W.~{Wegscheider}, and S.~{Ludwig}.
\newblock {Investigations on unconventional aspects in the quantum Hall regime
  of narrow gate defined channels}.
\newblock {\em Physica E Low-Dimensional Systems and Nanostructures},
  40:1130--1132, March 2008.

\bibitem{jose:prl}
A.~{Siddiki}, J.~{Horas}, J.~{Moser}, W.~{Wegscheider}, and S.~{Ludwig}.
\newblock {Interaction mediated asymmetries of the quantized Hall effect}.
\newblock {\em ArXiv e-prints:0905.0204[cond-mat.mes-hall]}, May 2009.

\bibitem{dassarma:mobility}
E.~H. {Hwang} and S.~{Das Sarma}.
\newblock {Limit to two-dimensional mobility in modulation-doped GaAs quantum
  structures: How to achieve a mobility of 100 million}.
\newblock {\em Phys. Rev. B}, 77(23):235437--+, June 2008.

\bibitem{Macleod09:background}
S.~J. {MacLeod}, K.~{Chan}, T.~P. {Martin}, A.~R. {Hamilton}, A.~{See}, A.~P.
  {Micolich}, M.~{Aagesen}, and P.~E. {Lindelof}.
\newblock {Role of background impurities in the single-particle relaxation
  lifetime of a two-dimensional electron gas}.
\newblock {\em Phys. Rev. B}, 80(3):035310--+, July 2009.

\bibitem{Guven03:115327}
K.~G{\"u}ven and R.~R. Gerhardts.
\newblock Self-consistent local-equilibrium model for density profile and
  distribution of dissipative currents in a hall bar under strong magnetic
  fields.
\newblock {\em Phys. Rev. B}, 67:115327, 2003.

\bibitem{siddiki2004}
A.~Siddiki and R.~R. Gerhardts.
\newblock Incompressible strips in dissipative hall bars as origin of quantized
  hall plateaus.
\newblock {\em Phys. Rev. B}, 70:195335, 2004.

\bibitem{Bilayersiddiki06:}
A.~Siddiki.
\newblock Self-consistent coulomb picture of an electron-electron bilayer
  system.
\newblock {\em Phys. Rev. B}, 75:155311, 2007.

\bibitem{Siddiki04:condmat}
A.~Siddiki and R.~R. Gerhardts.
\newblock The interrelation between incompressible srtips and quantized hall
  plateaus.
\newblock {\em Int. J. Mod. Phys. B}, 18:3541, 2004.

\bibitem{Gerhardts08:378}
R.~R. {Gerhartds}.
\newblock {The effect of screening on current distribution and conductance quantisation in narrow quantum Hall systems}.
\newblock {\em Phys. Stat. Sol.b}, 245:378, January 2008.

\bibitem{Siddiki:ijmp}
A.~Siddiki and R.~R. Gerhardts.
\newblock Range-dependent disorder effects on the plateau-widths calculated
  within the screening theory of the iqhe.
\newblock {\em Int. J. of Mod. Phys. B}, 21:1362, 2007.

\bibitem{Ando82:437}
T.~Ando, A.~B. Fowler, and F.~Stern.
\newblock Electronic properties of two-dimensional systems.
\newblock {\em Rev. Mod. Phys.}, 54:437, 1982.

\bibitem{Champel08:124302}
T.~{Champel}, S.~{Florens}, and L.~{Canet}.
\newblock {Microscopics of disordered two-dimensional electron gases under high
  magnetic fields: Equilibrium properties and dissipation in the hydrodynamic
  regime}.
\newblock {\em Phys. Rev. B}, 78(12):125302--+, September 2008.

\bibitem{TobiasK06:h}
T.~{Kramer}.
\newblock {a Heuristic Quantum Theory of the Integer Quantum Hall Effect}.
\newblock {\em International Journal of Modern Physics B}, 20:1243--1260, 2006.

\bibitem{Weichselbaum03:056707}
A.~{Weichselbaum} and S.~E. {Ulloa}.
\newblock {Potential landscapes and induced charges near metallic islands in
  three dimensions}.
\newblock {\em Phys. Rev E}, 68(5):056707--+, November 2003.

\bibitem{Sefa08:prb}
S.~{Arslan}, E.~{Cicek}, D.~{Eksi}, S.~{Aktas}, A.~{Weichselbaum}, and
  A.~{Siddiki}.
\newblock {Modeling of quantum point contacts in high magnetic fields and with
  current bias outside the linear response regime}.
\newblock {\em Phys. Rev. B}, 78(12):125423--+, September 2008.

\bibitem{Siddiki03:125315}
A.~Siddiki and R.~R. Gerhardts.
\newblock Thomas-fermi-poisson theory of screening for laterally confined and
  unconfined two-dimensional electron systems in strong magnetic fields.
\newblock {\em Phys. Rev. B}, 68:125315, 2003.

\bibitem{Lier94:7757}
K.~Lier and R.~R. Gerhardts.
\newblock Self-consistent calculation of edge channels in laterally confined
  two-dimensional electron systems.
\newblock {\em Phys. Rev. B}, 50:7757, 1994.

\bibitem{Oh97:13519}
J.~H. Oh and R.~R. Gerhardts.
\newblock Self-consistent thomas-fermi calculation of potential and current
  distributions in a two-dimensional hall bar geometry.
\newblock {\em Phys. Rev. B}, 56:13519, 1997.

\bibitem{Gerhardts75:285}
R.~R. Gerhardts.
\newblock Path-integral approach to the two-dimensional magneto-conductivity
  problem ii application ...
\newblock {\em Z. Physik B}, 21:285, 1975.

\bibitem{Laughlin81}
R.~B. Laughlin.
\newblock Gauge gedankenexperiment iqhe.
\newblock {\em Phys. Rev. B}, 23:5632, 1981.

\bibitem{Buettiker88:317}
M.~B{\"u}ttiker.
\newblock {\em IBM J. Res. Dev.}, 32:317, 1988.

\bibitem{Halperin82:2185}
B.~I. Halperin.
\newblock Self-consistent local-equilibrium model for density profile and
  distribution of dissipative currents in a hall bar under strong magnetic
  fields.
\newblock {\em Phys. Rev. B}, 25:2185, 1982.

\bibitem{Ahlswede01:562}
E.~Ahlswede, P.~Weitz, J.~Weis, K.~von Klitzing, and K.~Eberl.
\newblock Hall potential profiles in the quantum hall regime measured by a
  scanning force microscope.
\newblock {\em Physica B}, 298:562, 2001.

\bibitem{Yacoby00:3133}
S.~Ilani, A.~Yacoby, D.~Mahalu, and H.~Shtrikman.
\newblock Unexpected behavior of the local compressibility near the b=0
  metal-insulator transition.
\newblock {\em Phys. Rev. Lett.}, 84:3133, 2000.

\bibitem{Ashoori05:136804}
G.~A. {Steele}, R.~C. {Ashoori}, L.~N. {Pfeiffer}, and K.~W. {West}.
\newblock {Imaging Transport Resonances in the Quantum Hall Effect}.
\newblock {\em Physical Review Letters}, 95(13):136804--+, September 2005.

\bibitem{criticalexpo:09}
K.~{Slevin} and T.~{Ohtsuki}.
\newblock {Critical exponent for the quantum Hall transition}.
\newblock {\em Phys. Rev. B}, 80(4):041304(R), July 2009.

\bibitem{Serpilactivation:09}
S.~{Sakiroglu}, U.~{Erkarslan}, G.~{Oylumluoglu}, A.~{Siddiki}, and
  I.~{Sokmen}.
\newblock {Microscopic theory of the activated behavior of the quantized Hall
  effect}.
\newblock {\em ArXiv e-prints:0906.0661[cond-mat.mes-hall]}, June 2009.

\bibitem{Haug87:SdH}
R.~J. {Haug}, K.~V. {Klitzing}, and K.~{Ploog}.
\newblock {Analysis of the asymmetry in Shubnikov-de Haas oscillations of
  two-dimensional systems}.
\newblock {\em Phys. Rev. B}, 35:5933--5935, April 1987.

\bibitem{Siddiki02:Oxford}
A.~Siddiki and R.~R. Gerhardts.
\newblock Nonlinear thomas-fermi-poisson theory of screening for a hall bar
  under strong magnetic fields.
\newblock In A.~R. Long and J.~H. Davies, editors, {\em Proc. 15th Intern.
  Conf. on High Magnetic Fields in Semicond. Phys.}, Bristol, 2002. Institute
  of Physics Publishing.

\bibitem{Gerhardts87:asymmetry}
R.~J. {Haug}, R.~R. {Gerhardts}, K.~V. {Klitzing}, and K.~{Ploog}.
\newblock {Effect of repulsive and attractive scattering centers on the
  magnetotransport properties of a two-dimensional electron gas}.
\newblock {\em Physical Review Letters}, 59:1349--1352, September 1987.

\bibitem{Matthews05:497}
J.~{Matthews} and M.~E. {Cage}.
\newblock {Temperature Dependence of the Hall and Longitudinal Resistances in a
  Quantum Hall Resistance Standard}.
\newblock {\em J. Res. Natl. Inst. Stand. Technol.}, 110:497, October 2005.

\bibitem{afifPHYSEspin}
A.~{Siddiki}.
\newblock {The spin-split incompressible edge states within empirical
  Hartree approximation at intermediately large Hall samples}.
\newblock {\em Physica E Low-Dimensional Systems and Nanostructures},
  40:1124--1126, March 2008.

\end{thebibliography}
\bibliographystyle{unsrt}

\end{document}